\documentclass{article}
\usepackage[utf8]{inputenc}
\usepackage{amsmath}
\usepackage[rgb,dvipsnames]{xcolor}
\usepackage{tikz}
\usetikzlibrary {arrows.meta,bending,positioning} 
\usepackage{standalone}
\usepackage{bm}
\usepackage[sort&compress,square,numbers]{natbib}
\usepackage{graphicx}
\usepackage{amssymb}
\usepackage{amsthm}
\usepackage{hyperref}
\usepackage[scale=0.8]{geometry}
\usepackage{cleveref}
\Crefformat{figure}{Figure #2#1#3}
\Crefrangeformat{equation}{(#3#1#4)-(#5#2#6)}
\Crefname{equation}{Equation}{Equations}

\usepackage{float}
\usepackage{subcaption}

\usepackage{hhline}
\usepackage{makecell}

\newcommand{\beq}{\begin{equation}\begin{aligned}}
\newcommand{\eeq}{\end{aligned}\end{equation}}
\newcommand{\norm}[1]{||#1||}

\newcommand{\fd}[2]{\frac{\mathrm{d} #1 }{\mathrm{d} #2}}
\newcommand{\fdd}[2]{\frac{\mathrm{d}^2 #1 }{\mathrm{d} #2^2}}
\newcommand{\pd}[2]{\frac{\partial #1 }{\partial #2}}
\newcommand{\pdd}[2]{\frac{\partial^2 #1 }{\partial #2^2}}

\title{Patterns in Time and Space from a Single Morphogen via Nonlinear Layering}
\author{N. Mahashri\footnote{Department of Mathematics, School of Advanced Sciences,
Vellore Institute of Technology, Vellore 632014}, Andrew L. Krause\footnote{Department of Mathematical Sciences, Durham University, Stockton Road, Durham DH1 3LE,
UK, \href{mailto:andrew.krause@durham.ac.uk}{andrew.krause@durham.ac.uk}}
\addtocounter{footnote}{-2}\addtocounter{Hfootnote}{-2}, M. Chandru\footnotemark\addtocounter{footnote}{1}\addtocounter{Hfootnote}{1}, and Thomas E. Woolley\footnote{Cardiff School of Mathematics, Cardiff University, Senghennydd Road, Cardiff CF24 4AG, UK}  }

\date{}

\begin{document}

\maketitle
\begin{abstract}
    Spatial and temporal pattern formation in reaction-diffusion systems is typically studied with two or more equations, as scalar reaction-diffusion equations confined to convex domains do not admit stable inhomogeneous states in time or space on long timescales. Here, we show that a single morphogen diffusing across layered two-dimensional media, with nonlinear coupling between layers, is able to generate stable patterns in time and space. This $N$-layer model is analysed via a thin-domain limit, which reduces to an $N$-component reaction-diffusion system on a homogeneous one-dimensional domain. This reduced model can be analysed via linear stability techniques, showing that non-diffusive, or reactive, coupling between regions is necessary for pattern-forming instabilities, at least in the reduced model. This reduced system can exhibit Turing, Hopf, and Turing-wave instabilities, with emergent structures that are numerically shown to persist even away from the thin-domain regime of the full 2D single-morphogen system. These results suggest that heterogeneous stratification and nonlinear coupling can broaden the class of systems which exhibit complex spatiotemporal behaviours, which may be relevant in scenarios where only a single morphogen is known to act. 
\end{abstract}
\section{Introduction}

Minimal models of complex behaviours serve as valuable motifs for understanding wide classes of dynamics across a variety of biological and physical media. Two-species reaction-diffusion systems, for example, are some of the simplest models that exhibit spatially-periodic spontaneous pattern formation. Since Turing first proposed such models for biological pattern formation \citep{turing_chemical_1952}, numerous studies have analysed chemical and biological pattern forming systems in the search for mechanisms implicit in the simplest reaction-diffusion models, such as a long-range inhibitor and a short-range activator \citep{Maini2012}. Similarly, there has been tremendous work in characterising minimal models of oscillators \citep{Schnakenberg-1979-SCR, novak2008design}, with the smallest examples in continuous state and time being pairs of ordinary differential equations. Here we demonstrate that a single reacting and diffusing species with spatially-varying layers can, in the presence of nonlinearly-interacting boundary conditions, generate spatially and temporally complex states arising from linear instabilities analogous to multi-component systems (i.e.~Turing and Hopf instabilities).

Minimal models are powerful precisely because they trade mechanistic and biochemical specificity for conceptual generality. By reducing a system to the smallest set of ingredients capable of generating a phenomenon, they help identify which assumptions are essential, which are incidental, and which conclusions should persist across physical and biological realisations. On the other end of the complexity spectrum, detailed models play a complementary role. They are often better suited to calibration against data, quantitative prediction, and system-specific interpretation, but their conclusions are correspondingly tied more closely to a particular organism, geometry, parameter set, or experimental context. In this sense, minimal and complex models work in tandem to generate transferable organising principles and exploit them, respectively \citep{Servedio2014}.

Reaction-transport processes in layered, compartmentalised, and bulk-surface media are increasingly well-studied, as microscale geometry and interfacial exchange can qualitatively alter the effective large-scale dynamics \citep{stratified, diez2024turing}. In such settings, diffusion in one region may be coupled to reactions in another, so that pattern selection is shaped not only by local kinetics and transport coefficients, but also by how information is transmitted across interfaces and between spatial scales. This view has proven useful in stratified reaction-diffusion systems, where transverse coupling can significantly modify classical instability criteria. More generally, these geometrically-coupled models have been extensively explored in the context of bulk-surface and cell-bulk models, where lower-dimensional domains interact with higher-dimensional surroundings to generate new pattern-forming behaviour in a variety of contexts \citep{ratz_symmetry_2014,ratz_turingtype_2015,levine_membranebound_2005,madzvamuse_stability_2015,gomez_pattern_2021,paquin-lefebvre_pattern_2019,paquin-lefebvre_pattern_2020}. Recently, a simplified ODE coupled to the boundary of a scalar 1D reaction-diffusion equation has been rigorously shown to exhibit Hopf bifurcations and stable periodic oscillations \cite{pelz2026oscillations}. In some sense, this is a minimal model of a single-morphogen oscillator, if the nonlinear boundary condition can be interpreted as a self-interaction. 

Other lines of research within the scope of reaction-transport systems aim to broaden the class of models which exhibit complex spatiotemporal behaviours while maintaining some level of minimality. Curvature-inducing proteins have been theorised to provide a single-morphogen mechanochemical framework for short-range-activation and long-range inhibition \cite{mercker2013mechanochemical, veerman2021beyond, weevers2025mechanochemical, nesenberend2025curvature}. Reaction-diffusion models involving non-diffusible morphogens generalise the class of systems capable of producing stationary and spatiotemporal patterns from classical reaction-diffusion frameworks \cite{klika2012influence, korvasova2015investigating, kowall2025nonlinear}. See also the use of a single diffusible morphogen to initiate patterning in lattice-like models \cite{wang2022periodic}. Beyond reaction-diffusion formalisms, models of phase-separation are experiencing a renewed interest in biology as an alternative paradigm to Turing-like pattern formation \cite{boeynaems2018protein, menou2023physical}. Finally, we remark that spatial heterogeneity within reaction-diffusion systems can both modulate existing pattern-forming dynamics \cite{iron2001spike,page2003pattern, page2005complex, Krause2020WKBJ, woolley2021bespoke, gaffney_spatial_2023, krause2025pattern}, and generate novel behaviours beyond what their homogeneous counterparts can do  \cite{krause2018heterogeneity, kolokolnikov2018pattern, patterson2023spatial}.

As mentioned, some models are too simple to generate oscillatory dynamics in time or space. Specifically, scalar reaction-diffusion models on convex domains (with no flux conditions at the boundaries) are unable to generate stable oscillatory or spatially-inhomogeneous long-time solutions \citep{casten1978instability, matano1979asymptotic}. Although we remark that bistability can lead to very long-lived transients \cite{BronsardKohn} and, moreover, bistable dynamics in sufficiently nonconvex domains, such as the Matano dumbbell \cite{matano1983pattern}, can lead to stable inhomogeneous solutions. Our goal is to show, similar to Turing's original counterintuitive idea, that coupling such scalar non-patterning components across two-dimensional layered spatial regions can lead to spatio-temporal complexity.
Explicitly, via linear stability analysis of an asymptotically reduced model, we will demonstrate that a single morphogen interacting with itself across different layered regions is able to generate Turing, Hopf, and Turing-wave instabilities,  which we subsequently numerically verify in the full layered model. 

Critically, although our model is intentionally minimal, it also provides a multiscale framework in which vertical heterogeneity is resolved explicitly as layers, while a thin-layer limit converts that structure into an effective reaction-diffusion system on a horizontal line. In this reduction, the nonlinear interfacial fluxes are not merely auxiliary boundary conditions but the mechanism by which local information is transmitted, producing emergent macroscopic dynamics. This interpretation is also consistent with recent thermodynamically grounded approaches to coupled bulk-surface transport, where constitutive assumptions at interfaces determine the effective evolution at larger scales \cite{duda2023modelling}.

We present the model and an asymptotic reduction of it in \Cref{Sec:Modelling}. We discuss linear stability analysis of the full and reduced models in \Cref{Sec:Linear_Stability}. Examples of a variety of dynamical behaviours are given in \Cref{Sec:Examples}, with motivations for nonlinearities and parameters determined by the asymptotically reduced model. In \Cref{Sec:Conclusion} we conclude with key limitations of our approach and thoughts on important future directions.

\section{General model and asymptotic simplifications}\label{Sec:Modelling}

\begin{figure}
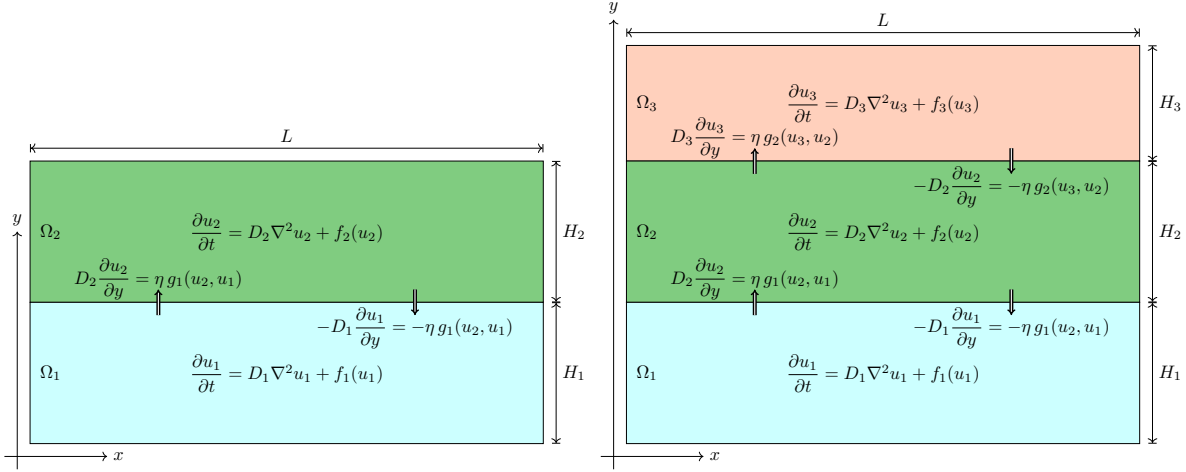

    \centering
    \includegraphics[width=0.45\linewidth]{N2_diagram.tex}
    \includegraphics[width=0.45\linewidth]{N3_diagram.tex}
    \caption{A diagram illustrating the model given by equations \Cref{main_multilayer_eq,neumann_boundary_conditions,coupling_boundary_conditions} for $N=2$ and $N=3$ layers.}
    \label{N=3_Diagram}
\end{figure}

Consider $N$ rectangular regions, $\Omega_i$, of size $L \times H_i$ for $i=1,\dots,N$, which are arranged vertically, as shown in \Cref{N=3_Diagram} for $N=2$ and 3. We pose the following general PDE for a scalar variable $u_i(t,x,y)$ on each region:
\beq\label{main_multilayer_eq}
    \pd{u_i}{t} = D_i \nabla^2 u_i + f_i(u_i), \quad x \in [0,L], \quad y \in [\hat{H}_{i-1},\hat{H}_i],
\eeq
where $\hat{H}_i = \sum_{j=1}^{i}H_j$, with $\hat{H_0}:=0$.  $\hat{H}_N$ is the height of all regions in total, allowing for a global coordinate system in $(x,y)$ despite distinct spatial domains for each equation. The left and right boundaries are given homogeneous Neumann conditions, as are the top and bottom boundaries. Explicitly,
\beq\label{neumann_boundary_conditions}
    \pd{u_i}{x}(t,0,y)=\pd{u_i}{x}(t,L,y)=0, \quad \pd{u_1}{y}(t,x,0)=\pd{u_N}{y}(t,x,\hat{H}_N)=0.
\eeq
We impose general (possibly nonlinear) boundary conditions between each internal layer of the form,
\beq\label{coupling_boundary_conditions}
    D_i\pd{u_i}{y}(t,x,\hat{H}_i)=\eta\,{g_{i}}({u}_{i+1}(t,x,\hat{H}_i),{u}_i(t,x,\hat{H}_i)) = D_{i+1}\pd{u_{i+1}}{y}(t,x,\hat{H}_i),
\eeq
where $\eta$ is a coupling constant between the layers. In the case that the $g_i$ are linear differences corresponding to diffusive fluxes (i.e. $g_i = u_{i+1}-u_i$), then $\eta$ has previously been interpreted as a permeability between the layers \citep{stratified}.

These boundary conditions conserve flux between regions, only redistributing mass despite being nonlinear. To see this, we denote $M_i$ as the total mass of $u_i$ in region $\Omega_i$. Integrating over $\Omega_i$ and $\Omega_{i+1}$ we have that
\beq
    &\pd{M_i}{t}+\pd{M_{i+1}}{t} = \iint_{\Omega_i}\pd{u_i}{t}dxdy+\iint_{\Omega_{i+1}}\pd{u_{i+1}}{t}dxdy \\
    =& \iint_{\Omega_i}D_i \nabla^2 u_i+f_i(u_i)dxdy+\iint_{\Omega_{i+1}}D_{i+1} \nabla^2 u_{i+1}+f_{i+1}(u_{i+1})dxdy \\
    =& \iint_{\Omega_i}f_i(u_i)dxdy+\iint_{\Omega_{i+1}}f_{i+1}(u_{i+1})dxdy +\int_0^LD_{i+1}\pd{u_{i+1}}{y}(t,x,\hat{H}_{i+1})-  D_i\pd{u_i}{y}(t,x,\hat{H}_{i-1})dx \\
    =& \iint_{\Omega_i}f_i(u_i)dxdy+\iint_{\Omega_{i+1}}f_{i+1}(u_{i+1})dxdy + \eta\int_0^L\left({g_{i+1}}({u}_{i+2},{u}_{i+1})|_{y=\hat{H}_{i+1}}-{g_{i-1}}({u}_{i},{u}_{i-1})|_{y=\hat{H}_{i-1}} \right)dx.
\eeq
We note that the first two terms on the right hand side of the last equality are the change of mass from reactions, and the last two terms are the contributions to the change of mass from the lower boundary of $\Omega_i$ and the upper boundary of $\Omega_{i+1}$, as the fluxes from the mutual boundary exactly cancel. If we perform the same summation over all regions $\Omega_i$, we would see that all of the internal coupling terms involving $g_i$ cancel out. Hence, these nonlinear fluxes generalize linear diffusion in terms of merely redistributing mass without creating or destroying it. Such nonlinear interfaces could then represent biochemical signal transduction across a cell membrane, for example.


\subsection{Thin-layer asymptotics}
For most choices of the functions $f_i$ and $g_i$, it is difficult to analytically study equations \Cref{main_multilayer_eq,neumann_boundary_conditions,coupling_boundary_conditions}. This is true even from the perspective of linear stability analysis, which we demonstrate in the next Section. Instead, inspired by the analysis of layered heterogeneous reaction-diffusion systems in Section 3.1 of \cite{fussell2019hybrid}, we will consider an asymptotic simplification which does allow straightforward analysis in thin domains. 

Throughout the following derivation we assume that the solutions and nonlinearities are sufficiently smooth for the formal manipulations below to be valid and that each layer has the same height $H_i=H$. We also impose the scaling $\eta=H\hat{\eta}$ with $\hat{\eta}=\mathrm{O}(1)$ as $H\to 0$, so that the interfacial coupling remains at leading order in the reduced model. Integrating in $y$ over the region $\Omega_i$, we find that
\beq\label{thin_asymptotics_eqn}
\int_{\hat{H}_{i-1}}^{\hat{H}_i} \pd{u_i}{t} dy &= \int_{\hat{H}_{i-1}}^{\hat{H}_i}D_i \nabla^2 u_i + f_i(u_i)dy \\
&= D_i\left( \pd{u_i}{y}(t,x,\hat{H}_i)- \pd{u_i}{y}(t,x,\hat{H}_{i-1})\right)+\int_{\hat{H}_{i-1}}^{\hat{H}_i} D_i \pdd{u_i}{x}+f_i(u_i)dy \\
&=\eta\left({g_{i}}({u}_{i+1},{u}_{i})-{g_{i-1}}({u}_{i},{u}_{i-1}) \right)+\int_{\hat{H}_{i-1}}^{\hat{H}_i} D_i \pdd{u_i}{x}+f_i(u_i)dy, 
\eeq
where we have omitted the dependencies on the $\hat{H}_i$ variables inside of the $g_i$ and $g_{i+1}$ functions, as these will reduce to the same point in the end.

We recall the following elementary limit:
\beq\label{limit_result}
    \lim_{H\to 0}\frac{1}{H}\int_{\hat{H}_{i-1}}^{\hat{H}_i}G(y)dy=\lim_{H\to 0}\int_{(i-1)}^{i}G(Hu)du = \int_{(i-1)}^{i}G(0)du= G(0),
\eeq
where we substituted $y = Hu$, and required that $G(y)$ is continuous across the interfaces. Now we divide \Cref{thin_asymptotics_eqn} by $H$ and take the limit $H \to 0$ to arrive at,
\beq\label{reduced_eqn}
    \pd{u_i}{t} = \hat{\eta}\left({g_{i}}({u}_{i+1},{u}_{i})-{g_{i-1}}({u}_{i},{u}_{i-1})\right) +D_i \pdd{u_i}{x}+f_i(u_i).
\eeq
Equation \eqref{reduced_eqn} also includes $i=1$ and $i=N$ if we set $g_0=g_{N}=0$, with the derivation following essentially the same reasoning. Hence, \Cref{reduced_eqn} describes $N$ coupled partial differential equations, which depend only on $t$ and $x$ on the same one-dimensional spatial domain $x \in [0,L]$. We remark that this analysis is somewhat crude, neglecting possible implications of boundary layers or issues in interchanging limiting processes. In particular, we will see that for nonlinear $g_i$, we do not generally anticipate continuity across interfaces, and so a more rigorous approach would need to resolve these as boundary layers near the interfaces. Nevertheless, we will show that this crude approximation gives a way to deduce qualitatively distinct dynamical regimes in the full 2D model, in a way that a direct analysis of the 2D model cannot easily do. 

\section{Linear instability analysis}\label{Sec:Linear_Stability}
We first study the general linear stability problem in 2D, subject to conditions on the existence of an equilibrium which is locally spatially homogeneous in each region. We deduce a dispersion relation for the growth of linear perturbations in terms of a rather complicated solvability condition. However, analysing this dispersion relation is nontrivial, even numerically, so we then perform an analysis of the reduced 1D model, which is much more amenable to general study.

\subsection{Linear stability analysis of the 2D problem}
We first sketch a linear stability analysis of the full 2D model given by system \Cref{main_multilayer_eq,neumann_boundary_conditions,coupling_boundary_conditions} to demonstrate its complexity. This has been done in the case where multiple species exist in each region, but when $g$ is a linear diffusive flux and there is a shared spatially-uniform equilibrium concentration, i.e.~$u_i = u^*$ such that $f_i(u^*)=g_i(u^*,u^*)=0$ \citep{stratified, diez2024turing}. This previous analysis led to a complicated transcendental dispersion relation, which was, in general, not easy to use, and was analysed in asymptotic regimes, or numerically. We will find a similar result here, using the result primarily as motivation for the thin-domain asymptotics introduced before.

In our model we do not expect a shared equilibrium between each region (as this would require fine-tuning of the functions $f_i$ and $g_i$), or even one where the solutions are spatially homogeneous within each region. In the absence of spatially homogeneous equilibria, there are some possible regimes where a Turing-like analysis can be carried out \cite{Krause2020WKBJ, gaffney_spatial_2023}, though the analysis becomes rather involved even in 1D domains with simple boundary conditions. Hence we consider the situation where there is a constant equilibrium solution, $u_i^*$, in each region that is compatible with the boundary conditions, i.e.~$f_i(u_i^*)=0$ and $g_i(u_{i+1}^*,u_i^*)=0$. 

We perturb this equilibrium via $u_i = u_i^* + \varepsilon U_i(t,x,y)$ assuming $\varepsilon \ll 1$ to arrive at the linearised system,
\beq
    \pd{U_i}{t} = D_i \nabla^2 U_i + f_i'(u_i^*)U_i, \quad x \in [0,L], \quad y \in [\hat{H}_{i-1},\hat{H}_i],
\eeq
where the $U_i$ satisfy the no-flux boundary conditions \eqref{neumann_boundary_conditions}, as well as the following linearised internal conditions,
\beq\label{linearized_coupling_conditions}
    D_i\pd{U_i}{y}(t,x,\hat{H}_i)=\eta\left(\pd{g_i}{u_{i+1}}(u_{i+1}^*,u_i^*)U_{i+1}(t,x,\hat{H}_i)+\pd{g_i}{u_{i}}(u_{i+1}^*,u_i^*)U_{i}(t,x,\hat{H}_i) \right) = D_{i+1}\pd{U_{i+1}}{y}(t,x,\hat{H}_i).
\eeq

We assume separable solutions of the form $U_i = e^{\lambda t}\cos(k_q x)Y_i(y)$, where $k_q = q\pi/L$ for integer $q$. Substituting this into the above linearized system, we have that perturbation growth rates are determined by,
\beq
    \lambda Y_i =  -D_ik_q^2Y_i + D_iY_i'' + f_i'(u_i^*)Y_i, \quad y \in [\hat{H}_{i-1},\hat{H}_i],
\eeq
which can be rearranged to find the following second-order differential equations for $Y_i$,
\beq\label{Y_i_eqns}
    \fdd{Y_i}{y} \equiv {Y_i''} = c_i^2Y_i, \textrm{ where } c_i = \sqrt{\frac{\lambda-f_i'(u_i^*)}{D_i} +k_q^2},  \quad y \in [\hat{H}_{i-1},\hat{H}_i],
\eeq
with the associated boundary conditions,
\beq\label{Y_i_boundaries}
    D_iY_i'(\hat{H}_i)=\eta\left(a_iY_{i+1}(\hat{H}_i)+d_iY_{i}(\hat{H}_i) \right) = D_{i+1}Y_{i+1}'(\hat{H}_i), \quad Y_1'(0) =0, \quad Y_N'(\hat{H}_N)=0,
\eeq
where
$$
a_i = \pd{g_i}{u_{i+1}}(u_{i+1}^*,u_i^*), \quad d_i = \pd{g_i}{u_{i}}(u_{i+1}^*,u_i^*).
$$

Equations \eqref{Y_i_eqns}-\eqref{Y_i_boundaries} are a system of $N$ oscillator-like equations for the $Y_i$, along with $2(N-1)$ coupling boundary conditions and 2 homogeneous Neumann conditions arising from the outermost layers. We can solve this system by supposing solutions of the form\footnote{Implicitly, we are assuming real $\lambda$ with $\lambda > 0$ to motivate the use of hyperbolic functions; one could work with general exponential functions for complex $c_i$ instead, arriving at similarly complicated relationships in the end.},
\beq
Y_i = \gamma_i\cosh(c_iy) + \beta_i\sinh(c_iy),
\eeq
leading to internal coupling conditions of the form, 
\beq
&D_ic_i\left(\gamma_i\sinh(c_i \hat{H}_i) + \beta_i\cosh(c_i\hat{H}_i) \right) = D_{i+1}c_{i+1}\left(\gamma_{i+1}\sinh(c_{i+1} \hat{H}_i) + \beta_{i+1}\cosh(c_{i+1}\hat{H}_i) \right)\\=& \eta \left (a_i(\gamma_{i+1}\cosh(c_{i+1} \hat{H}_i)+\beta_{i+1}\sinh(c_{i+1} \hat{H}_i)) + d_i(\gamma_i\cosh(c_i\hat{H}_i) + \beta_i\sinh(c_i\hat{H}_i)) \right).\label{consistency}
\eeq
The condition  $Y'(0)=0$ fixes $\beta_1=0$, and the condition at $y=\hat{H}_N$ requires $\gamma_N\sinh(c_N \hat{H}_N)+\beta_N\cosh(c_N \hat{H}_N) =0.$

In total, with $\beta_1=0$, equation \eqref{consistency} leads to $2N-1$ linear equations relating the $2N-1$ nontrivial $\gamma_i$ and $\beta_i$ to one another that must be solved. These equations can be written as a homogeneous linear system, so a solvability condition (determinant) can be used to determine a condition for nontrivial solutions in $\gamma_i$ and $\beta_i$. However, this solvability condition is a complicated transcendental relationship, as in Equation (28) of \citep{stratified}, which must be solved numerically, or asymptotically, for the growth rates $\lambda$ embedded in the coefficients $c_i$. We anticipate that, fixing all other parameters including $k_q$, there are a countably infinite number of solutions for $\lambda$ corresponding to increasingly oscillatory eigenfunctions, $Y_i(y)$. However, given that evaluating the roots of this equation is difficult even numerically, we instead opt to analyse the reduced model, \Cref{reduced_eqn}, directly as a system of $N$ coupled reaction-diffusion equations. 

\subsection{Analysis of the reduced model}
\Cref{reduced_eqn} is a standard $N$-component reaction-diffusion system. We remark that linear stability analysis for $N$-species reaction-diffusion equations has been studied by many authors \citep{satnoianu2000turing, villar2023general, villar2025designing}, particularly focusing on Turing and Turing-wave instabilities leading to spatial and spatiotemporal pattern formation. We note that there are a variety of block-matrix results applicable to models like \Cref{reduced_eqn}, where each layer contains $M$ species, and the $g_i$ are linear diffusive couplings \citep{catlla2012instabilities}. As our system is scalar with nonlinear couplings, we will not use these results but instead proceed directly to demonstrate the key elements of what can drive instabilities in this system.

We again begin by assuming that there is an equilibrium of \Cref{reduced_eqn}, i.e.~that there exists real numbers $u_i^*$ such that $f_i(u_i^*)=g_i(u_{i+1}^*,u_i^*)-g_{i-1}(u_{i}^*,u_{i-1}^*)=0$ for all\footnote{Note that the assumptions on the equilibrium values $u_i^*$ are slightly more general in this case than in the 2D scenario above, where we required $g_i(u_{i+1}^*,u_i^*)=0$. This is possible as fluxes from one region may exactly balance fluxes from another, as there is no possibility of breaking spatial homogeneity in the vertical direction in the 1D model.} $i$. Linearizing about this state by setting $u_i = u_i^*+\varepsilon U_i(t,x)$, we find that perturbations satisfy,
\beq\label{linearized_reduced_eqn}
    \pd{U_i}{t} = \hat{\eta}\left(\pd{g_i}{u_{i+1}}U_{i+1}+\left(\pd{g_i}{u_{i}}-\pd{g_{i-1}}{u_i}\right)U_i-\pd{g_{i-1}}{u_{i-1}}U_{i-1}\right) +D_i \pdd{U_i}{x}+f_i'(u_i^*)U_i,
\eeq
where the derivatives of $g_i$ and $g_{i-1}$ are evaluated at $(u_{i+1}^*,u_i^*)$ and $(u_{i}^*,u_{i-1}^*)$ respectively. We now define the matrix,
\beq
    \bm{L} =
\begin{pmatrix}
b_1 & a_1 & 0 & 0 & \cdots & 0 & 0 \\
-b_1 & -a_1+b_2 & a_2 & 0 & \cdots & 0 & 0 \\
0 & \ddots & \ddots & \ddots & & \vdots & \vdots \\
0 & \cdots & -b_{i-1} & -a_{i-1}+b_i & a_i & \cdots & 0 \\
\vdots & & & \ddots & \ddots & \ddots & 0 \\
0 & 0 & \cdots & 0 & -b_{N-2} & -a_{N-2}+b_{N-1} & a_{N-1} \\
0 & 0 & \cdots & 0 & 0 & -b_{N-1} & -a_{N-1}
\end{pmatrix},
\eeq
where
\beq
a_i = \pd{g_i}{u_{i+1}}(u_{i+1}^*,u_i^*), \quad b_i = \pd{g_i}{u_{i}}(u_{i+1}^*,u_{i}^*).
\eeq
We also define the diagonal matrices $\bm{D}_{ij} = D_i\delta_{ij}$, and $\bm{J}_{ij} = f_i'(u_i^*)\delta_{ij}$, where $\delta_{ij}$ is the Kronecker $\delta$. Denoting $\bm{U}$ as the vector with components $U_i$, we can write \Cref{linearized_reduced_eqn} as,
\beq
    \pd{\bm{U}}{t} = \hat{\eta}\bm{L}\bm{U} + \bm{D} \pdd{\bm{U}}{x}+\bm{J}\bm{U}.
\eeq

We can study the stability of these perturbations by assuming that $\bm{U} \propto e^{\lambda t}\cos(k_q x)$, where $k_q$ is again an integer multiple of $\pi/L$ denoting the wavenumber in the $x$ direction. For each $k_q$, the growth rates $\lambda$ are then eigenvalues of the matrix
\beq\label{reduced_matrix}
    \bm{M}_{k_q} = \hat{\eta}\bm{L}-k_q^2\bm{D}+\bm{J}.
\eeq
We want to focus on systems which are stable in the absence of coupling and diffusion, and hence assume that $f_i'(u_i^*)\leq0$. This means that the terms given by $-k_q^2\bm{D}+\bm{J}$ constitute a negative-semidefinite diagonal matrix for any real $k_q$. So any instability must arise from the coupling with $\bm{L}$. 

We note that this does not necessarily mean that $\bm{L}$ itself has a positive eigenvalue. For example, in \citep{satnoianu2005some} it is demonstrated that an unstable matrix can arise from a stable diagonal matrix being added to another stable matrix, where by \emph{stable} we mean that all real parts of all eigenvalues are negative (semi)-definite. On the other hand, we can see by arguments from \citep{neubert2002transient} that if $\bm{M}_{k_q}$ has a positive real part eigenvalue, then $\bm{L}$ must at least be a `reactive' matrix in the sense that its Hermitian part, given by $H(\bm{L})=\left(\bm{L}+\bm{L}^T\right)/2$, has a positive eigenvalue (see \citep{neubert1997alternatives} for the definition of reactivity and its relationship with the largest eigenvalue of the Hermitian part of a matrix). 

We now demonstrate the need for reactivity of $\bm{L}$, following the arguments from \citep{neubert2002transient} which generalize to our setting. Reactivity is indicative either of an unstable linear system, or of non-exponential transient growth of perturbations to a linear system which is eventually exponentially damped. We therefore consider the more general perturbation ansatz $\bm{U} = \bm{v}(t)\cos(k_q x)$, whereby $\bm{v}$ now satisfies
\beq
    \fd{\bm{v}}{t} = \bm{M}_{k_q}\bm{v} = \hat{\eta}\bm{L}\bm{v} -k_q^2\bm{D}\bm{v}+\bm{J}\bm{v}.
\eeq
Letting $\lambda_1(\bm{M}_{k_q})$ denote the eigenvalue of $\bm{M}_{k_q}$ with largest real part then, using Rayleigh's principle \citep{horn2012matrix}, we have that
\beq
\fd{}{t}\norm{\bm{v}} = \fd{}{t} \sqrt{\bm{v}^T \bm{v}} = \frac{\bm{v}^T(\bm{M}_{k_q}+\bm{M}_{k_q}^T)\bm{v}}{2\norm{\bm{v}}} = \frac{\bm{v}^TH(\bm{M}_{k_q})\bm{v}}{\norm{\bm{v}}} \leq \lambda_1(H(\bm{M}_{k_q})).
\eeq
where
\beq
H(\bm{M}_{k_q}) =  \hat{\eta}H(\bm{L}) -k_q^2\bm{D}+\bm{J},
\eeq
due to $\bm{D}$ and $\bm{J}$ being diagonal. Further, since the Hermitian part is symmetric, it has only real eigenvalues.

If  $\lambda_1(H(\bm{M}_{k_q}))<0$, then linear instabilities are not possible. By Weyl's inequality \citep{horn2012matrix}, the largest eigenvalue of a sum of symmetric matrices is less than or equal to the sum of the largest eigenvalue of each matrix. Hence for instabilities to be possible, we need $\lambda_1(H(\bm{L}))>0$, implying that $\bm{L}$ must be a reactive matrix.

We can also rule out linear diffusive coupling as a possibility for driving Turing or Hopf instabilities. Consider $g_i = u_{i+1}-u_i$, i.e.~discrete diffusion between layers, which leads to the coupling term of the form $g_i(u_{i+1},u_i)-g_{i-1}(u_i,u_{i-1}) = u_{i+1}-2u_i+u_{i-1}$, which is the discrete Laplacian on a 1D lattice. The discrete Laplacian on any graph is negative-semidefinite, and in the case of the 1D lattice has eigenvalues given by $\lambda_j(\bm{L}) = 2\cos\left((j-1)\pi/N\right)-2$. As the Laplacian matrix is symmetric, again by Weyl's inequality it cannot induce instability when added to a negative diagonal matrix. More generally, asymmetric forms of $\bm{L}$ leading to some kind of reactivity as above are necessary to drive instabilities. 

We remark that other general claims can be made about the eigenvalues of $\bm{M}_{k_q}$, particularly using tools from systems theory and network science. For example, the form of \Cref{reduced_matrix} is essentially the same kind of matrix studied using master stability conditions (and functions) in \citep{porter2016dynamical}, where there is a local reaction term coupled nonlinearly via a graph structure. In our case, the underlying graph is a 1D lattice. Rather than dwell on the general case, however, we proceed to consider specific examples of instabilities in this simplified system that can be shown to persist in the full 2D model.

\section{Examples of linear instabilities}\label{Sec:Examples}
We now develop several example systems exhibiting different kinds of linear instabilities. These examples are developed by first considering reduced 1D models which can be easily analysed, and then simulating the corresponding 2D systems to demonstrate the persistence of spatially and/or temporally inhomogeneous solution behaviours. We detail the local kinetics and coupling functions of the systems we consider in \Cref{system_table}.

The full two-dimensional multilayer model equations \Cref{main_multilayer_eq,neumann_boundary_conditions,coupling_boundary_conditions} were solved using a finite element method implemented in MATLAB, which can be found at \citep{SMMLM2026code}. Triangular meshes were generated using the MATLAB PDE Toolbox, and the spatial discretisation employed continuous piecewise-linear finite elements. To ensure consistent spatial resolution across different geometries, the mesh spacing was defined as $dy = H/10$ and $dx$ proportional to $L$, maintaining a fixed number of elements across each layer thickness and consistent resolution along the domain length. For example, in the thin-layer regime $H=0.1$ of length $L=20$, this corresponds to $dx=0.08$ and $dy=0.01$, yielding approximately $3{,}254$ nodes and $5{,}986$ triangular elements per layer. A constant time step of $dt=0.001$ was used for all simulations. Diffusion terms were treated implicitly using a Crank–Nicolson discretisation, while reaction and interlayer coupling terms were treated explicitly within an implicit–explicit (IMEX) time integration scheme. Such finite element approaches are standard for reaction–diffusion systems and avoid the need for nonlinear solvers at each timestep \citep{thomee2007galerkin, hundsdorfer2013numerical}. Simulations were run until the solution reached a long-time regime, with convergence determined by the relative change between successive time steps falling below a tolerance of $10^{-9}$.

\begin{table}
    \centering
    \begin{tabular}{|c|c|c|c|c|c|}\hline
         $N$ & System & Kinetics
         & Coupling & 1D Equilibrium & 2D Equilibrium  \\ \hhline{|=|=|=|=|=|=|}
         2 & Autocatalytic & $f_1 = b, f_2 = a-u_2$
         & $g_1 = -u_2^2u_1$ &  $\left(\frac{b}{\hat{\eta}(a+b)^2},a+b\right)$ & None \\ \hline
         2 & In-phase & \makecell{$f_1=bu_1-\varepsilon u_1^3$,\\ $f_2=au_2-\varepsilon u_2^3$}
         & $g_1=(u_1+u_2)$ & (0,0) & (0,0) \\ \hline
         3 & Turing-wave & $f_i = p_i u_i-u_i^3$
         & \makecell{$g_1 = p_4u_1 + p_5u_2$,\\ $g_2 = p_6u_2+p_7u_3$} & $(0,0,0)$ & $(0,0,0)$ \\ \hline
    \end{tabular}
    \caption{A summary of the three systems that we consider, defined by their local kinetics $f_i(u_i)$ and coupling functions $g_i(u_{i+1}, u_i)$. We also include the spatially homogeneous equilibrium of the system which we focus our analysis on. In the 1D autocatalytic model this is a unique spatially homogeneous equilibrium, though there are additional nonzero equilibria in the other two systems.}
    \label{system_table}
\end{table}

We validated our numerical approach by comparing 2D simulations with outputs of the same model simulated in COMSOL Multiphysics \citep{Comsol-2021}. Similarly, simulations of the 1D reduced model, \Cref{reduced_eqn}, were carried out using both simple finite difference schemes in MATLAB, and using VisualPDE \citep{walker2023visualpde}. We also directly checked spatial and temporal convergence by refining the mesh resolution and time step, confirming that the pattern dynamics remained unchanged for sufficiently small step sizes. 

All simulations were initialised from the spatially homogeneous equilibrium with small-amplitude random perturbations and run for sufficiently long times to observe asymptotic behaviour. Throughout this Section, we will show 2D panel plots using the same aspect ratio, but vary $H$ and $L$. Hence, the reader is advised to pay close attention to the axes limits, as the underlying simulations are carried out for a wide variety of different aspect ratios. Finally, we remark that the parameters in the reduced model were used to find suitable parameters in the 2D model, with $\hat{\eta}$ often used to explore a range of behaviour. Unless otherwise reported, in 1D simulations we take $\hat{\eta}=1$. 

\subsection{Turing and Hopf instabilities in a Schnakenberg-like autocatalytic model}
We first consider the autocatalytic model given in \Cref{system_table}. The reduced 1D system, \Cref{reduced_eqn}, then appears as a rescaled version of the well-studied Schnakenberg system \citep{Schnakenberg-1979-SCR},
\beq \label{eq:reduced_schnak}
    \pd{u_1}{t} = D_1\pdd{u_1}{x} + b - \hat{\eta}u_2^2 u_1,
    \quad
    \pd{u_2}{t} = D_2\pdd{u_2}{x} + a - u_2 + \hat{\eta}u_2^2 u_1.
\eeq

Linear stability analysis of this reduced model to spatially homogeneous and inhomogeneous perturbations is very standard \citep{murray_mathematical_2003}. The homogeneous equilibrium is stable without diffusion if $\hat{\eta}(a+b)^3>b-a$, otherwise it exhibits a Hopf instability. Setting $D_2=1$ for notational simplicity, the necessary conditions for Turing instability (up to wavenumber selection requiring a sufficiently large domain) are then, 
\begin{equation}
    (a+b)^3<\frac{D_1}{\hat{\eta}}(b-a), 
    \quad 
    \left(\frac{D_1}{\hat{\eta}}(b-a)-(a+b)^3\right)^2>4\frac{D_1}{\hat{\eta}}(a+b)^4,
\end{equation}
which, for $D_1/\hat{\eta} \gg 1$, is satisfied for $b>a$, noting that we also need parameters to satisfy the homogeneous stability condition of $\hat{\eta}(a+b)^3>b-a$. We remark that this model exhibits both Turing and Hopf instabilities, as well as the codimension-2 bifurcation where these instabilities coincide, around which one can find a variety of spatiotemporal states as well as spatially homogeneous oscillations \citep{TuringHopfSchnakenberg}. As the system is of cross-kinetic type, linear theory predicts that any patterned solutions will have $u_1$ and $u_2$ being out-of-phase \citep{murray_mathematical_2003}.

The coupling matrix for this 1D system is given by
\beq
\hat{\eta}\bm{L} = \begin{pmatrix}
    -\hat{\eta} (a+b)^2 & -\frac{2b}{a+b}\\\hat{\eta} (a+b)^2 & \frac{2b}{a+b} 
\end{pmatrix}, 
\eeq
which always has one zero eigenvalue, and the other given by $\lambda = 2b/(a+b)-\hat{\eta}(a+b)^2$. Importantly, the Hermitian part always satisfies $\det(H(\bm{L}))<0$, indicating that the coupling matrix is always reactive even when $\bm{L}$ is itself stable. While this is not sufficient to predict when Turing or Hopf instabilities might arise, it gives some insight into the role of the coupling terms in driving instabilities.

\subsubsection*{Out-of-phase Turing patterns}

\begin{figure}
    \centering
    \begin{subfigure}[b]{0.45\textwidth}
        \includegraphics[width=\textwidth]{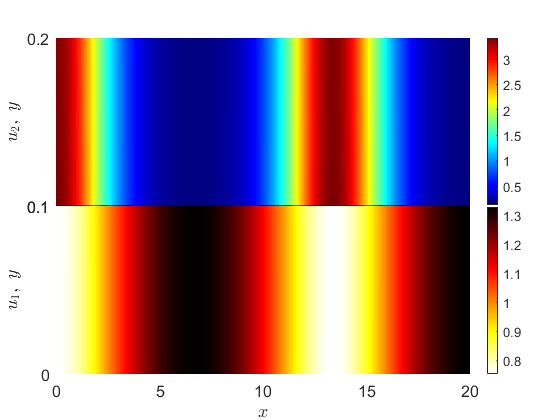}
        \caption{$H=\eta=0.1$, $t\approx250$}
        
    \end{subfigure}
    \begin{subfigure}[b]{0.45\textwidth}
        \includegraphics[width=\textwidth]{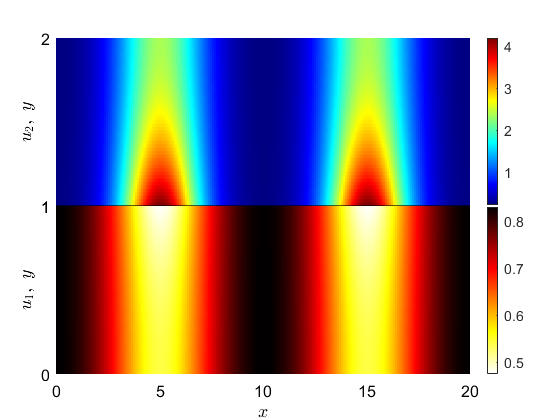}
        \caption{$H=\eta=1$, $t\approx250$}
        
    \end{subfigure}
    \begin{subfigure}[b]{0.45\textwidth}
        \includegraphics[width=\textwidth]{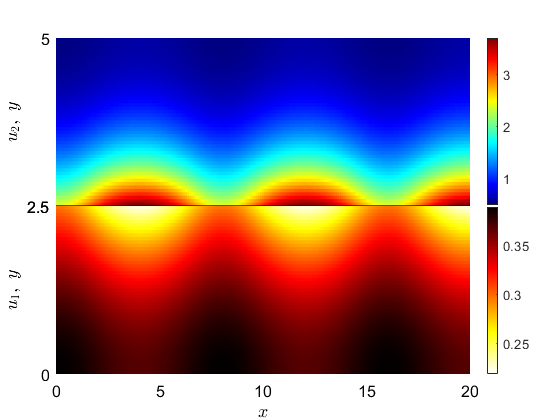}
        \caption{$H=\eta=2.5$, $t\approx500$}
        
    \end{subfigure}
    \begin{subfigure}[b]{0.45\textwidth}
        \includegraphics[width=\textwidth]{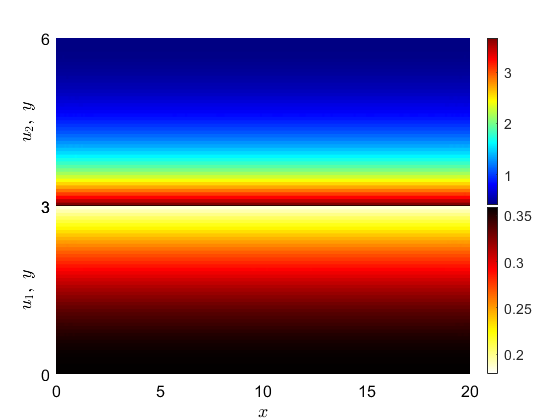}
        \caption{$H=\eta=3$, $t\approx100$}
        
    \end{subfigure}
\caption{Long-time solutions to equations \Cref{main_multilayer_eq,neumann_boundary_conditions,coupling_boundary_conditions} for the autocatalytic model given in \Cref{system_table}.  The parameters are given by $a=0.05$, $b=1.2$, $D_1=1$, $D_2=30$, and $L=20$.}
\label{fig:turing_outofphase}
\end{figure}

\Cref{fig:turing_outofphase} illustrates simulations in the Turing regime of the 2D autocatalytic model (see \Cref{system_table}), characterised by spatial maxima in one layer coinciding with minima in the other. For small $H=\eta$, both layers develop stationary stripe patterns that are nearly uniform across the vertical direction, with amplitudes and wavelengths consistent with the dynamics of the thin-layer reduction given by \Cref{eq:reduced_schnak} (with $\hat{\eta} = 1$). 

As the layer height, $H$, and coupling strength, $\eta$, increase, the spatial organisation undergoes a qualitative transition. Vertical gradients grow in magnitude, and concentration extrema begin to concentrate near the interfacial boundary. This indicates that vertical transport, neglected at leading order in the asymptotic reduction, becomes important. For sufficiently large values, e.g.~$H=\eta=3$, the horizontal periodicity is suppressed entirely. As seen in \Cref{fig:turing_outofphase}(d), the solution converges to a horizontally stratified steady state with no variation along $x$, although a significant gradient is maintained across the layer thickness due to the implicit heterogeneity in the system along the $y$ direction. This transition highlights one limitation of the thin-domain approximation, indicated in \Cref{system_table}. Namely, the 2D model does not admit a spatially-homogeneous equilibrium, so outside of the pattern-forming regime, the system tends towards a heterogeneous equilibrium which cannot be analysed through the methods in \Cref{Sec:Linear_Stability}. While the 1D reduced model predicts persistent Turing patterns for these parameter values, the full two-dimensional geometry shows that vertical diffusion can modulate the interfacial coupling to effectively stabilise the system against horizontal symmetry-breaking towards a vertically varying equilibrium.

\subsubsection*{Hopf oscillations}

\begin{figure}
    \centering
    \begin{subfigure}[b]{0.45\textwidth}
        \includegraphics[width=\textwidth]{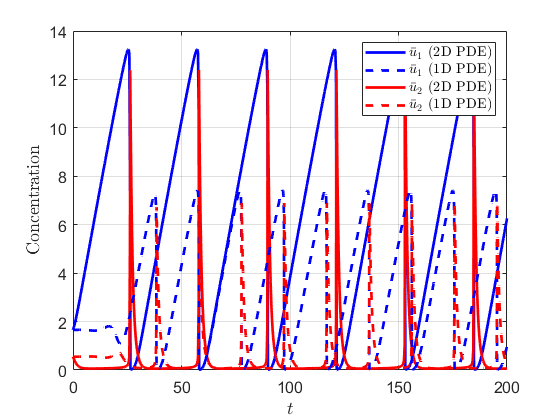}
        \caption{$b=0.5$}
        
    \end{subfigure}
    \begin{subfigure}[b]{0.45\textwidth}
        \includegraphics[width=\textwidth]{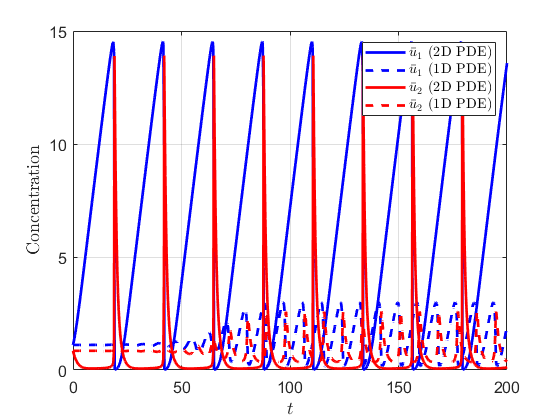}
        \caption{$b=0.8$}
        
    \end{subfigure}
    \begin{subfigure}[b]{0.45\textwidth}
        \includegraphics[width=\textwidth]{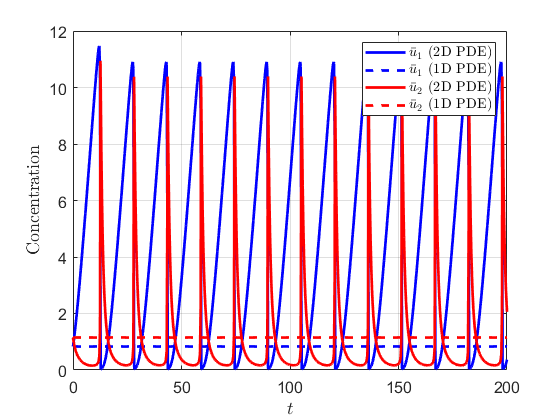}
        \caption{$b=1.1$}
        
    \end{subfigure}
    \begin{subfigure}[b]{0.45\textwidth}
        \includegraphics[width=\textwidth]{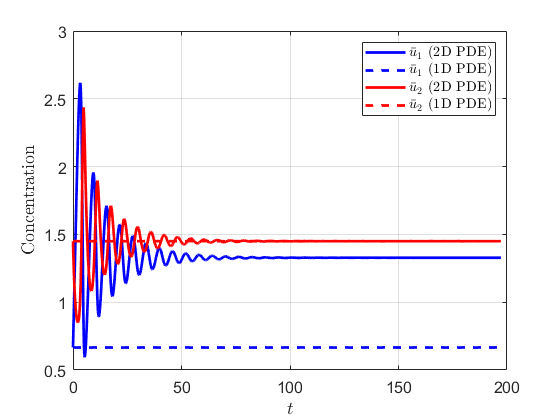}
        \caption{$b=1.4$}
        
    \end{subfigure}
\caption{Temporal dynamics comparing equations \Cref{main_multilayer_eq,neumann_boundary_conditions,coupling_boundary_conditions} for the autocatalytic model given in \Cref{system_table} (solid lines) with the corresponding one-dimensional reduction in~\Cref{eq:reduced_schnak} (dashed lines). The parameters are given by $a=0.05$, $D_1=D_2=1$, $H=\eta=0.1$, and $L=20$. In both models, we plot the spatial averages $\bar{u}_i$, given by $\bar{u}_i=\frac{1}{L}\int_0^L u_i(x,t)\,dx$ for the 1D PDE and by $\bar{u}_i=\frac{1}{|\Omega_i|}\int_{\Omega_i}u_i(x,y,t)\,dx\,dy$ for the 2D PDE.}
\label{fig:hopf}
\end{figure}

\begin{figure}
    \centering
    \begin{subfigure}[b]{1\textwidth}
    \centering
        \includegraphics[width=0.45\textwidth, height=5.7cm]{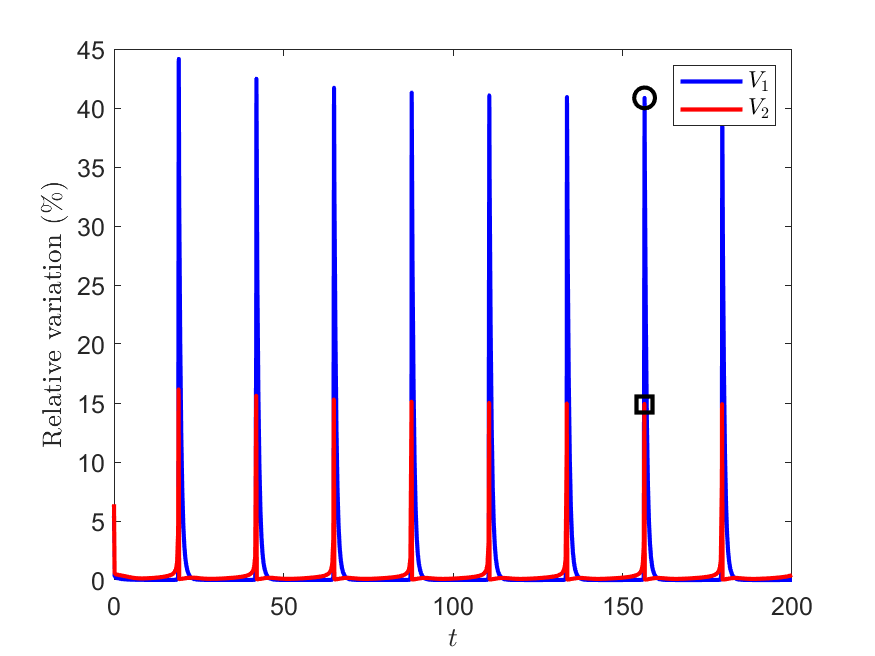}
        \caption{}
    \end{subfigure}
    \begin{subfigure}[b]{0.45\textwidth}
        \includegraphics[width=\textwidth]{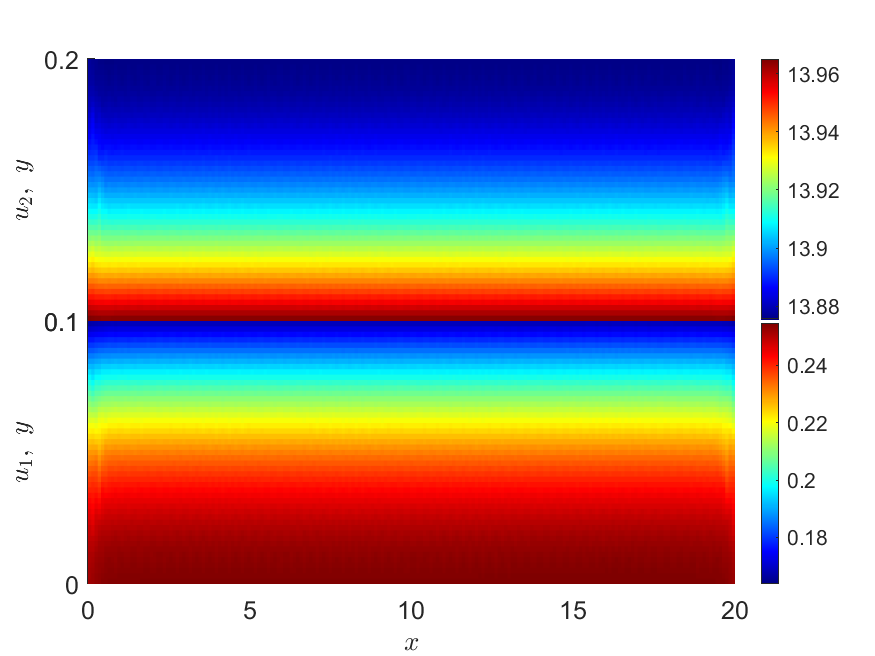}
        \caption{$t=156.518$}
    \end{subfigure}
    \begin{subfigure}[b]{0.45\textwidth}
        \includegraphics[width=\textwidth]{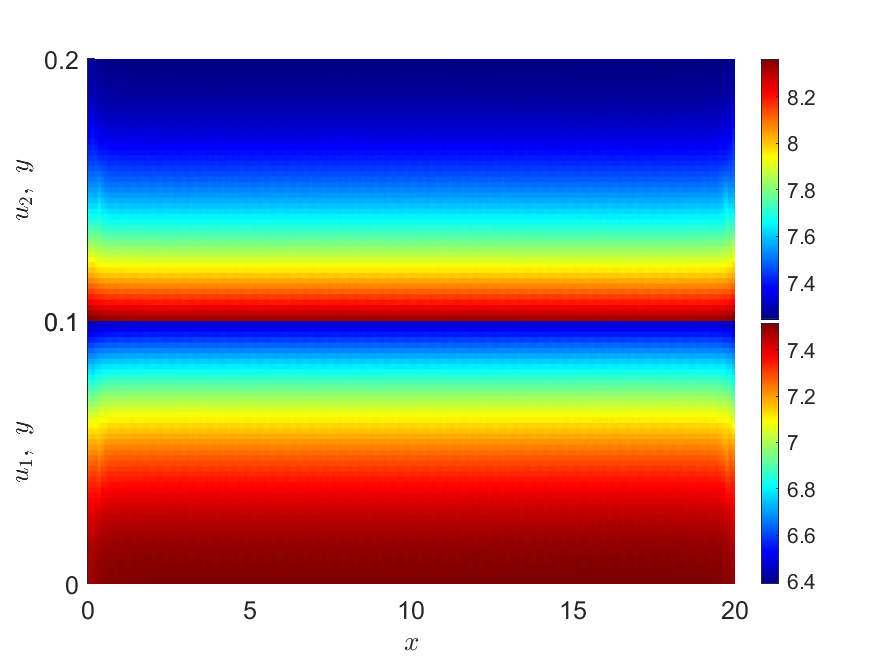}
        \caption{$t=156.462$}
    \end{subfigure}
\caption{Two-dimensional Hopf dynamics for $b=0.8$. (a) Relative variation in each layer over time, given by $V_i(t)=100\,(\max u_i-\min u_i)/\mathrm{mean}(u_i)$, with the circle- and square-marked peaks corresponding to the selected times shown in the snapshots (b) and (c), illustrating pronounced vertical variation during the oscillation cycle.}
\label{fig:hopf_2D}
\end{figure}

We next examine oscillatory instabilities by fixing $a = 0.05$ and $D_1 = D_2 = 1$, and varying the parameter $b$ across representative values crossing the Hopf bifurcation threshold $\hat{\eta}(a+b)^3=b-a$ of the reduced model. To ensure consistency between the full system and the reduced model, we set $H = \eta = 0.1$, so that $\hat{\eta} = 1$ is fixed in the reduced model. 

\Cref{eq:reduced_schnak} admits a spatially homogeneous limit cycle when $D_1=D_2$, which is also reflected in simulations of the corresponding one-dimensional reduced PDE system, up to possible phase shifts arising from spatial transients. For different values of $b$, we compare spatial averages $\bar u_i$ computed from both the one-dimensional reduced PDE system and the full 2D model, as shown in \Cref{fig:hopf}.

For smaller values of $b$ (e.g.\ $b = 0.5$ and $b = 0.8$), both models exhibit sustained oscillations consistent with a limit cycle arising from a Hopf bifurcation. As $b$ increases, the oscillation amplitude decreases, and for sufficiently large $b$ (e.g.\ $b = 1.4$), both models approach a stable homogeneous steady state. Thus, we see that the reduced model captures the qualitative transition between oscillatory and non-oscillatory behaviour. However, clear quantitative discrepancies are observed between the two models. In particular, differences arise in oscillation amplitudes, waveform shapes, and the location of the effective equilibrium around which oscillations occur. These differences persist even for smaller values of $H = \eta$, indicating that the thin-layer reduction does not fully capture the detailed dynamics in this regime.

As illustrated in \Cref{fig:hopf_2D}(a), the solution remains nearly homogeneous for most of the oscillation cycle, with short-lived spikes in vertical variation occurring during rapid phases of the oscillation, leading to sharp gradients in $y$. The snapshots in \Cref{fig:hopf_2D}(b,c) show the corresponding transient vertical gradients arising at two representative spikes in this variation. This indicates that the reduced model captures the onset of temporal behaviour but does not fully resolve spatial structure in the full system, as it neglects all vertical variation. These sharp gradients undoubtedly contribute to the discrepancy in period, amplitude, and onset parameters of these oscillations between the 1D and 2D models.

\subsubsection*{Boundary regime near the Turing-Hopf transition}

\begin{figure}
    \centering
    \begin{subfigure}[b]{0.45\textwidth}
        \includegraphics[width=\textwidth]{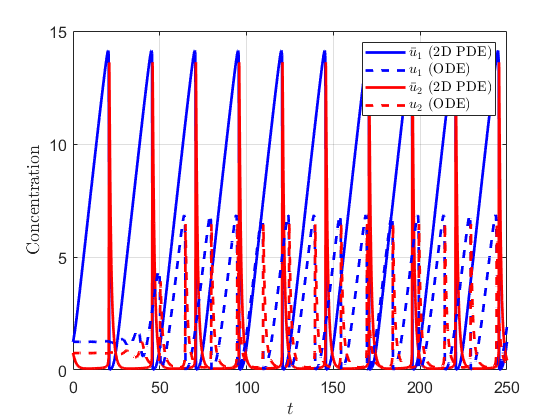}
        \caption{$\eta=0.1$ ($\hat{\eta}=1$ in the 1D Model)}
        
    \end{subfigure}
    \begin{subfigure}[b]{0.45\textwidth}
        \includegraphics[width=\textwidth]{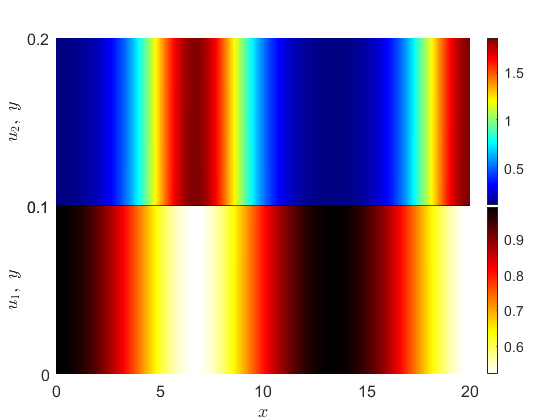}
        \caption{$\eta=0.25$, $t\approx200$}
        
    \end{subfigure}
    \begin{subfigure}[b]{0.45\textwidth}
        \includegraphics[width=\textwidth]{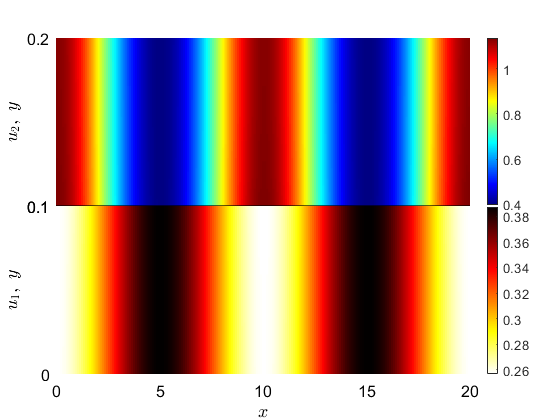}
        \caption{$\eta=0.75$, $t\approx200$}
        
    \end{subfigure}
    \begin{subfigure}[b]{0.45\textwidth}
        \includegraphics[width=\textwidth]{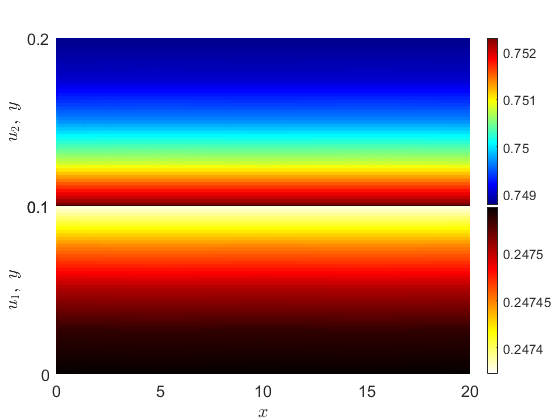}
        \caption{$\eta=1$, $t\approx500$}
        
    \end{subfigure}
\caption{Boundary-induced dynamics of equations \Cref{main_multilayer_eq,neumann_boundary_conditions,coupling_boundary_conditions} for the autocatalytic model given in \Cref{system_table}. The parameters are given by $a=0.05$, $b=0.7$, $D_1=1$, $D_2=20$, $H=0.1$, and $L=20$. (a) Temporal evolution comparing the spatial average of 2D PDE (solid line) with the corresponding ODE system (dashed line). (b)-(d) Long-time two-dimensional solutions of the full model.}
\label{fig:boundary}
\end{figure}


In \Cref{fig:boundary}, we explore transitions between the distinct dynamical regimes of the Schnakenberg model by varying the coupling strength $\eta$ while maintaining a fixed domain height $H=0.1$, in a parameter regime close to the boundary between Hopf and Turing instabilities predicted by the reduced model. This parameter sweep reveals a sequence of qualitative changes in the observed dynamics associated with increasing the strength of the interlayer transport.

At low coupling strength ($\eta=0.1$), the system exhibits temporal oscillations, as shown in \Cref{fig:boundary}(a), which displays oscillatory behaviour similar to that observed in the Hopf regime in \Cref{fig:hopf}, with nearly-homogeneous oscillations outside of short-lived sharp gradients. The dashed curve in this panel represents the spatially homogeneous (ODE) dynamics of the reduced model subject to spatially-homogeneous perturbations. Although the concentration varies periodically in time, the solution remains uniform along the $x$-direction, while exhibiting a noticeable variation in the vertical ($y$) direction. We note that the corresponding one-dimensional reduced PDE does not exhibit sustained oscillations in this regime, instead converging to a patterned state with amplitude comparable to \Cref{fig:boundary}(b). Such sensitivity to initial data and parameters is well-known in Turing-Hopf regimes \cite{kuznetsov2017pattern, sanchez2019turing}.

For intermediate coupling values, \Cref{fig:boundary}(b,c) demonstrates that the temporal oscillations are replaced by stationary stripe patterns along the $x$-direction, indicating the emergence of a diffusion-driven spatial instability in the full two-dimensional geometry. This is consistent with a shift in the dominant instability from a spatially homogeneous (Hopf-type) mode to a finite-wavenumber (Turing-type) mode as the coupling strength is increased.

When the coupling is increased further to $\eta=1$ in \Cref{fig:boundary}(d), longitudinal patterning is again suppressed, and the solution relaxes toward a vertically stratified steady state with no variation along the $x$-direction. Together, these results show that varying the coupling strength alone can move the system between oscillatory, stationary patterned, and spatially homogeneous regimes in the full geometry, underscoring the influence of transverse transport and domain structure on the resulting dynamics. As in the Hopf regime, the reduced model captures the range of dynamical behaviours observed in the full system; however, differences arise in the parameter values at which transitions between these regimes occur.

\subsection{In-phase Turing patterns} 


We next consider a model which exhibits in-phase Turing instabilities (sometimes known as being a `pure activator-inhibitor system'). The classical example of such a system is the model proposed by Gierer and Meinhardt \citep{gierer_theory_1972}, which cannot be represented in terms of the functions in \Cref{reduced_eqn}. Instead, we consider the in-phase model in \Cref{system_table}. This has the reduced 1D form:
\beq\label{eq:reduced_lin_cubic}
    \pd{u_1}{t} &= D_{1}  \pdd{u_1}{x} + b u_1-\varepsilon {u_1}^{3}+\hat{\eta} \left(u_1+u_2\right),\\
    \pd{u_2}{t} &= D_{2}  \pdd{u_2}{x} + a u_2 -\varepsilon {u_2}^{3}-\hat{\eta} \left(u_1+u_2\right),
\eeq
where $\varepsilon$ is a small parameter needed for instabilities to saturate. We consider the $(u_1^*,u_2^*) =(0,0)$ steady state which is stable in the absence of diffusion as long as $a+b<0$ and $\hat{\eta}(b-a) < ab$. For $\hat{\eta}>0$, this then implies $b<0$. We will again set $D_2=1$ for simplicity. This 1D system can exhibit a Turing instability if $D_1a +b>0$ and $(D_1a +b)^2-4D_1(ab+\hat{\eta}(b-a))>0$. These conditions then necessitate that $a>0$, so that $D_1>1$ is necessary for patterning. We note that the coupling matrix has a double zero eigenvalue, but its Hermitian part has eigenvalues $\pm\hat{\eta}$.

The 1D model also exhibits steady states $(u_1^*,u_2^*) = \left(0,\pm \sqrt{a/\varepsilon}\right)$, as well as mixed steady states which bifurcate from these as $\hat{\eta}$ increases from zero. Hence, we focus on cases where $\varepsilon$ is sufficiently small so that these steady states are far away from $(0,0)$, which we want to exhibit Turing-like patterns. For an interactive exploration of this 1D model with these parameters, we refer to the VisualPDE simulation\footnote{\url{https://visualpde.com/sim/?mini=4uoESYdL}. This uses the website described in \citep{walker2023visualpde} for real-time interactive PDE simulations.}, where, by varying $\hat{\eta}$, one can observe the transition between homogeneity for $\hat{\eta}\gtrapprox 0.6$, to regions of periodic patterns for $0.6 \gtrapprox \hat{\eta} \gtrapprox 0.3$, and finally to bistable patterns only evident in $u_2$ for $\hat{\eta}\lessapprox 0.3$. Such bistable `patterns' are precisely those observed in the scalar Allen-Cahn equation for the $u_2$ system in isolation, and on long timescales for $\hat{\eta}=0$ will eventually coarsen to a single front or a homogeneous state over very long timescales \citep{BronsardKohn}.

\begin{figure}[h!!!t!!!b!!!p]
    \centering
    \begin{subfigure}[b]{0.32\textwidth}
        \includegraphics[width=\textwidth]{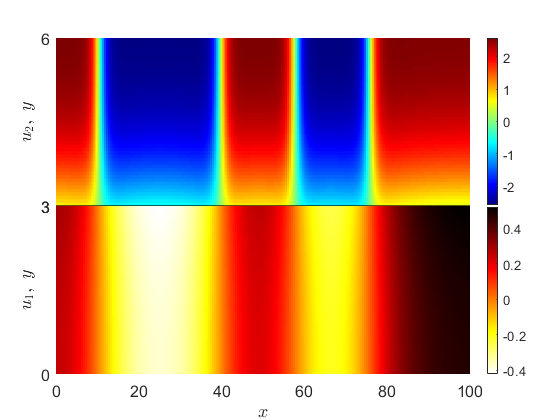}
        \caption{$H=\eta=3$}
        
    \end{subfigure}
    \begin{subfigure}[b]{0.32\textwidth}
        \includegraphics[width=\textwidth]{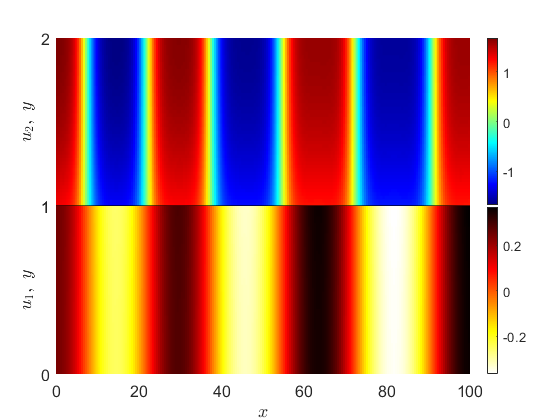}
        \caption{$H=\eta=1$}
        
    \end{subfigure}
    \begin{subfigure}[b]{0.32\textwidth}
        \includegraphics[width=\textwidth]{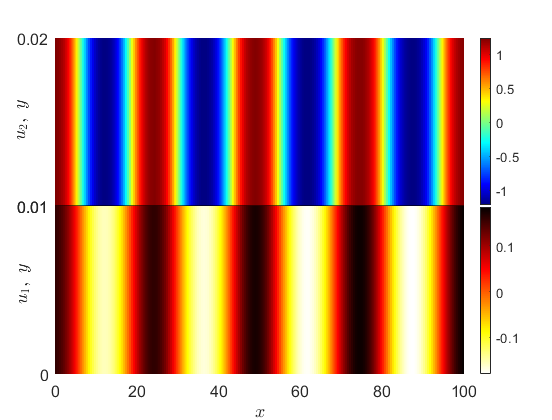}
        \caption{$H=\eta=0.01$}
        
    \end{subfigure}\\

    \begin{subfigure}[b]{0.33\textwidth}
        \includegraphics[width=\textwidth]{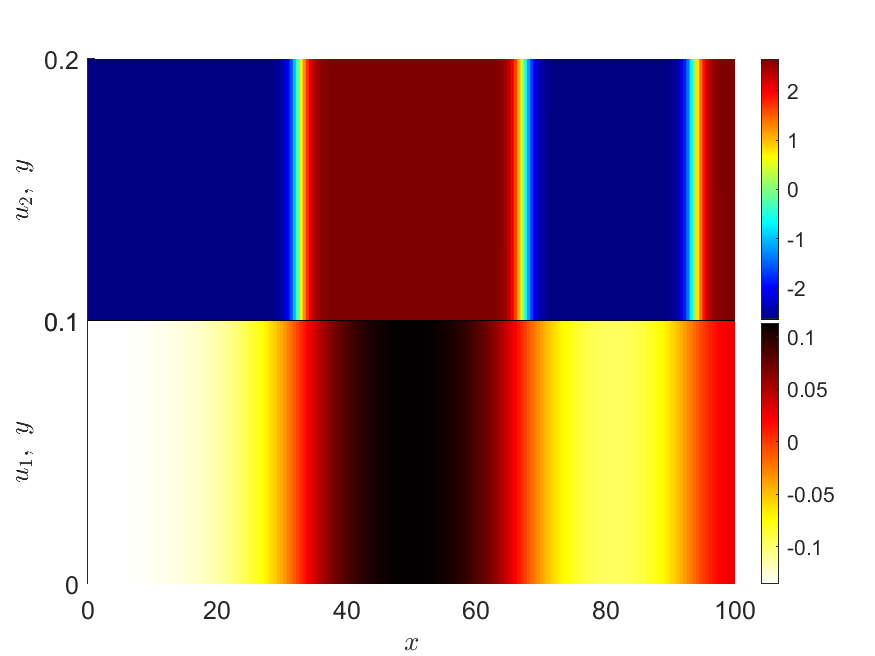}
        \caption{$H=0.1$, $\eta=0.01$}
        
    \end{subfigure}
\caption{Stationary spatial structures of equations~\Cref{main_multilayer_eq,neumann_boundary_conditions,coupling_boundary_conditions} for the `In-phase' model given in \Cref{system_table}. The parameters are given by $a=0.75$, $b=-1$, $\varepsilon=0.1$, $D_1=50$, and $D_2=1$. Solutions are displayed at $t=250$.}
\label{fig:in_phase}
\end{figure}

\Cref{fig:in_phase} illustrates these dynamics for the linear-cubic kinetics given by \Cref{eq:reduced_lin_cubic}. Panels (a) and (d) exhibit qualitatively different behaviour to (b) and (c). Specifically, in (a) and (d) the solution amplitude approaches the bistable equilibria $u_2=\pm\sqrt{a/\varepsilon}$, and the spatial structures resemble domains separated by metastable fronts rather than periodic patterns. Such dynamics are characteristic of bistable phase separation rather than periodic patterning due to a diffusion-driven instability. Panel (a) corresponds to the case $H=\eta$ with large coupling and layer height, while panel (d) represents the weak-coupling regime $(\eta \ll H)$, where the $u_2$ layers behave nearly independently, and $u_1$ is only loosely mirroring the dynamics in the upper domain.

Panels (b) and (c) display spatially periodic structures whose amplitudes remain well below the bistable equilibrium magnitude $\sqrt{a/\varepsilon}\approx 2.74$. These patterns do not correspond to transitions between the bistable states of the cubic terms but instead arise from a small-amplitude diffusion-driven instability. Consistent with this interpretation, the spatial structure depends on the diffusion contrast between the layers.

Repeating the simulations with equal diffusion coefficients $D_1=D_2$ shows behaviour that depends strongly on the value of $H=\eta$. For smaller and moderate values (e.g.\ $H=\eta=0.01,\, 1$), the solution relaxes to the homogeneous steady state $u_i=0$, indicating that the nonlinear coupling at the interface dominates and suppresses spatial structure. In contrast, for larger values (e.g.\ $H = \eta = 3$), the homogeneous state becomes unstable and the system develops spatially heterogeneous patterns comparable to \Cref{fig:in_phase}(a,d). This behaviour is consistent with the presence of positive linear growth away from the interface (via the $a u_2$ term), which can destabilise the zero state when the domain is sufficiently large, leading to bistable pattern formation. 

Taken together, these results suggest that spatially periodic patterns occur only within an intermediate coupling regime, as was seen in the 1D reduced model. When the coupling is too weak $(\eta \ll H)$, the layers evolve almost independently, and the bistable kinetics dominate. For sufficiently strong coupling ($H=\eta$ large), the dynamics again approach large-amplitude domain structures associated with the bistable equilibria. Between these limits, diffusion-driven instabilities produce the small-amplitude periodic patterns observed in panels (b)--(c). We do note that for $a>0$, the zero equilibrium is never stable in the 2D model, as perturbations away from the interface will always grow. Hence there is some subtlety in differentiating these regimes. 

\subsection{Turing-wave instabilities}

A Turing-wave (sometimes `oscillatory Turing') instability is one where the linear growth rates are stable for $k_q=0$ but for some $k_q>0$ there is a complex-conjugate pair of growth rates with $\Re(\lambda)>0$. As an example of this, we use the methods detailed in \citep{villar2025designing} to design a simple $N=3$-component system which admits these instabilities. We use the kinetics and coupling detailed in \Cref{system_table} under Turing-wave to implement this system. This gives the  reduced model,
\beq \label{eq:reduced_wave_model}
     \pd{u_{1}}{t} &= D_1 \pdd{u_{1}}{x} + p_{1} u_{1}-{u_{1}}^{3}+ \hat{\eta} \left(p_{4} u_{1} + p_{5} u_{2}\right),\\
      \pd{u_{2}}{t} &= D_2\pdd{u_{2}}{x} + p_{2} u_{2}-{u_{2}}^{3}+ \hat{\eta} \left(-p_{4} u_{1} - p_{5} u_{2}+p_{6} u_{2} + p_{7} u_{3}\right),\\
      \pd{u_{3}}{t} &= D_3 \pdd{u_{3}}{x} + p_{3} u_{3}-{u_{3}}^{3}+ \hat{\eta}\left(-p_{6} u_{2} - p_{7} u_{3}\right).
\eeq
Here the equilibrium of interest is given by $u_i^*=0$. Rather than analyse this model in great detail (which is rather involved), we remark that parameter sets can be found such that this equilibrium is stable in the absence of diffusion and coupling (i.e.~$p_1<0,p_2<0,p_3<0$), but for which the full system exhibits a Turing-wave instability. For an interactive exploration of this 1D model with these parameters, we refer to the VisualPDE simulation\footnote{\url{https://visualpde.com/sim/?mini=5tCqd8fe}.}.

Here, the coupling matrix is given by
\beq \label{eq:coupling_matrix_wave}
\bm{L} = \begin{pmatrix}
    p_4 & p_5 & 0\\
    -p_4 & -p_5+p_6 & p_7\\
    0 & -p_6 & -p_7
\end{pmatrix}.
\eeq
This matrix has one zero eigenvalue (as the columns each sum to 0), and two other eigenvalues. For our parameters, these other two eigenvalues are given by $\lambda \approx 2 \pm 4.89i$ (to two decimal places) and hence the wave instability is directly generated via the coupling matrix.

\begin{figure}
\begin{minipage}{0.92\textwidth}
    \begin{subfigure}[b]{0.33\textwidth}
        \includegraphics[width=\textwidth]{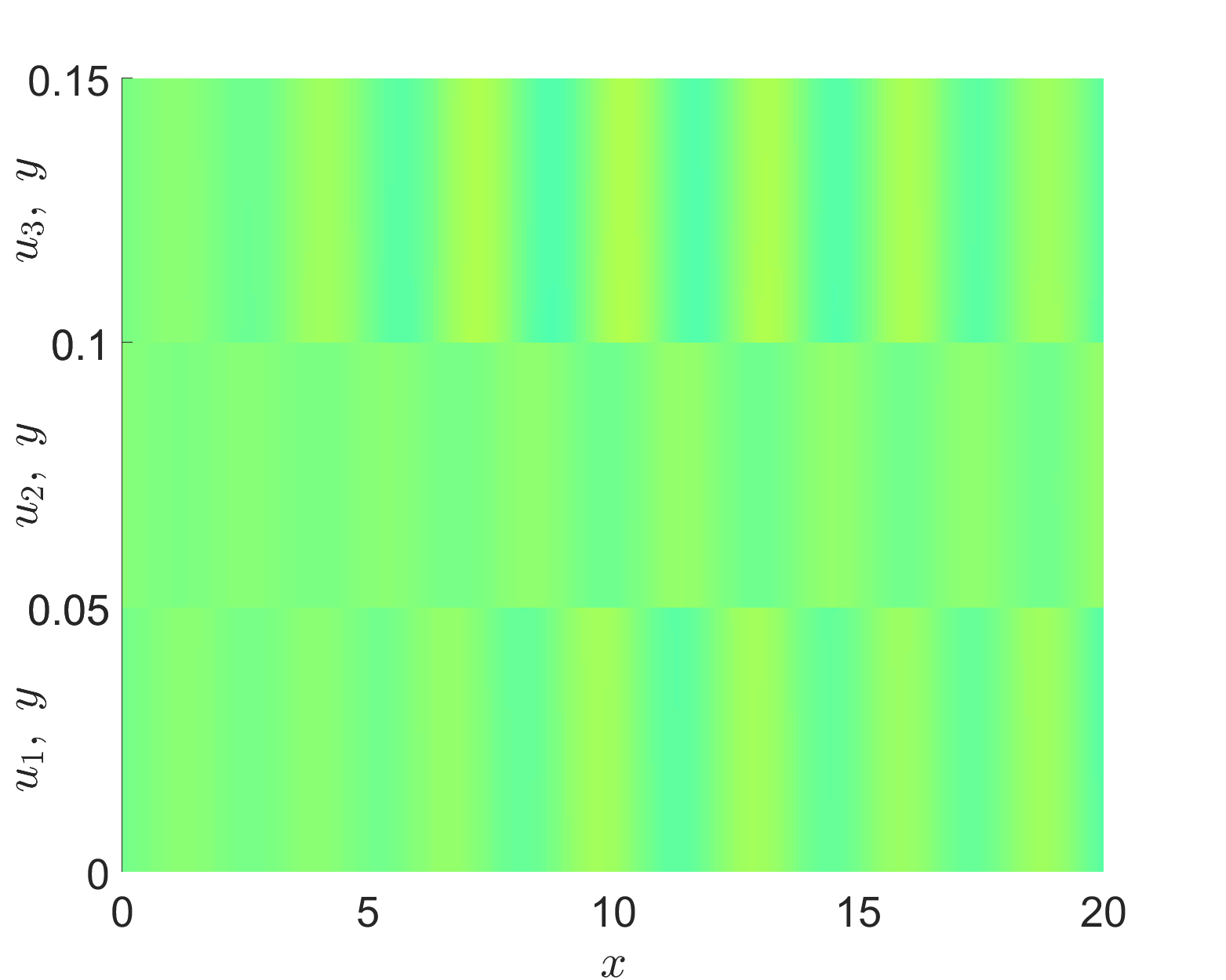}
        \caption{$t=10$}
        
    \end{subfigure}
    \begin{subfigure}[b]{0.33\textwidth}
        \includegraphics[width=\textwidth]{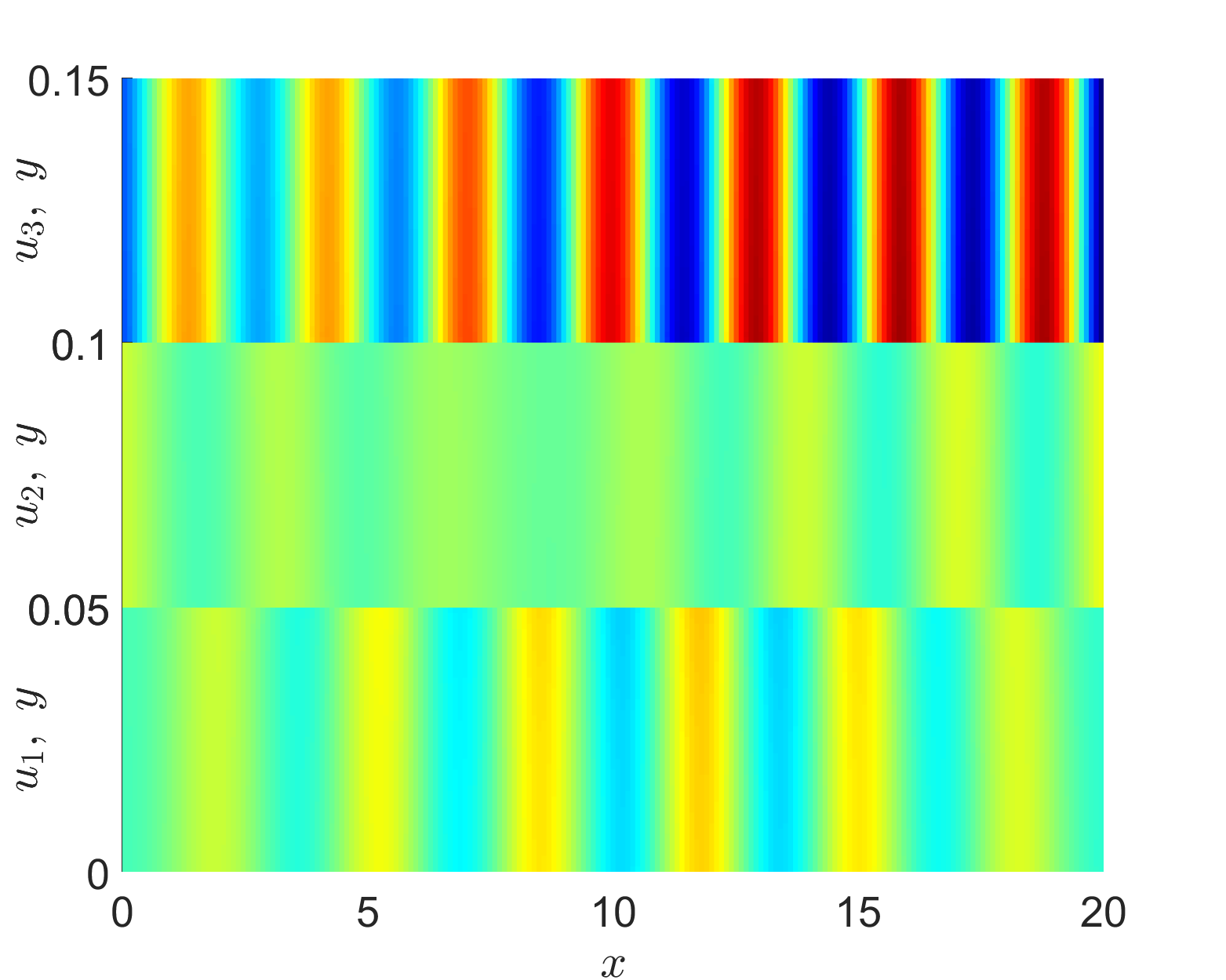}
        \caption{$t=15$}
        
    \end{subfigure}
    \begin{subfigure}[b]{0.33\textwidth}
        \includegraphics[width=\textwidth]{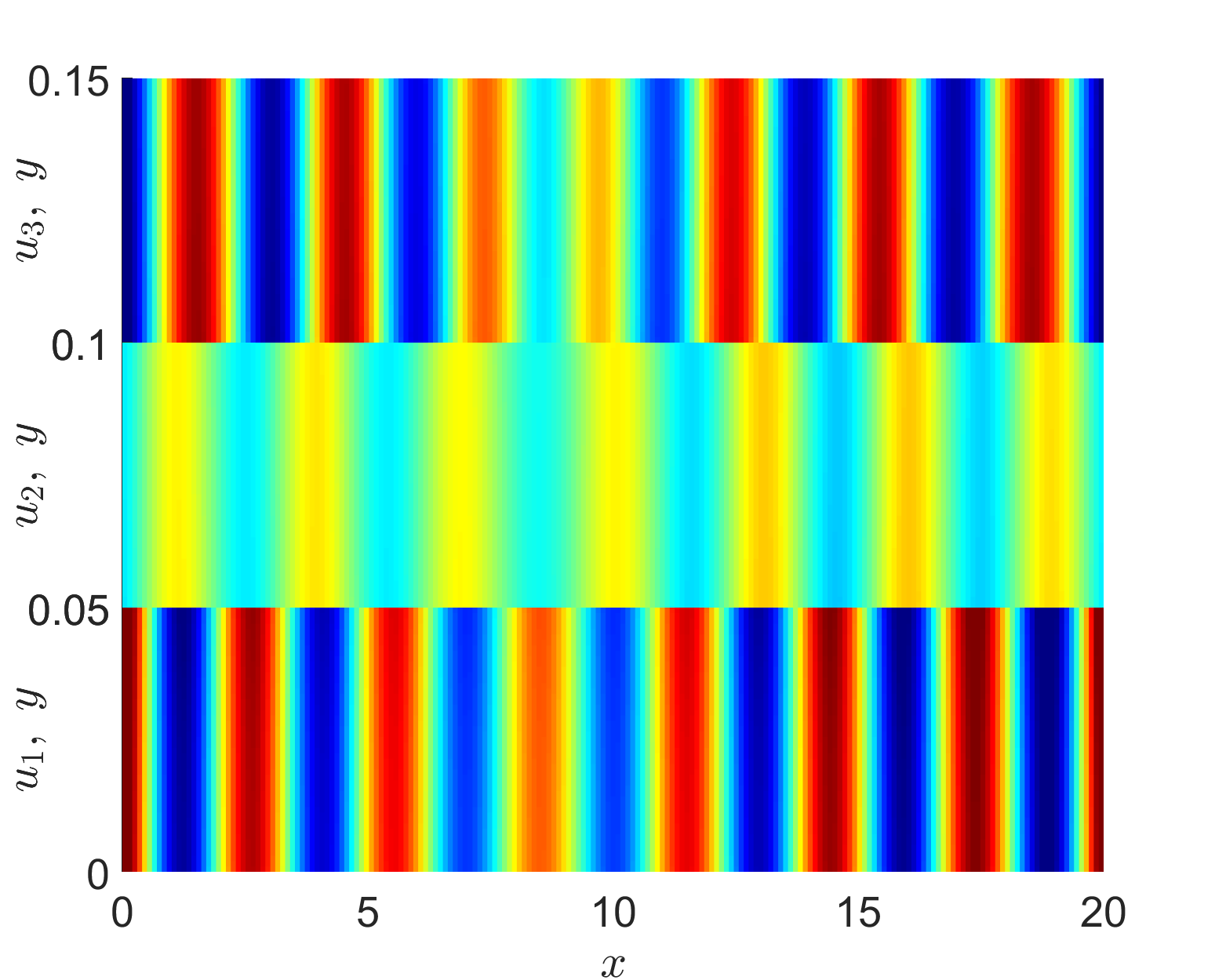}
        \caption{$t=20$}
        
    \end{subfigure}
    \begin{subfigure}[b]{0.33\textwidth}
        \includegraphics[width=\textwidth]{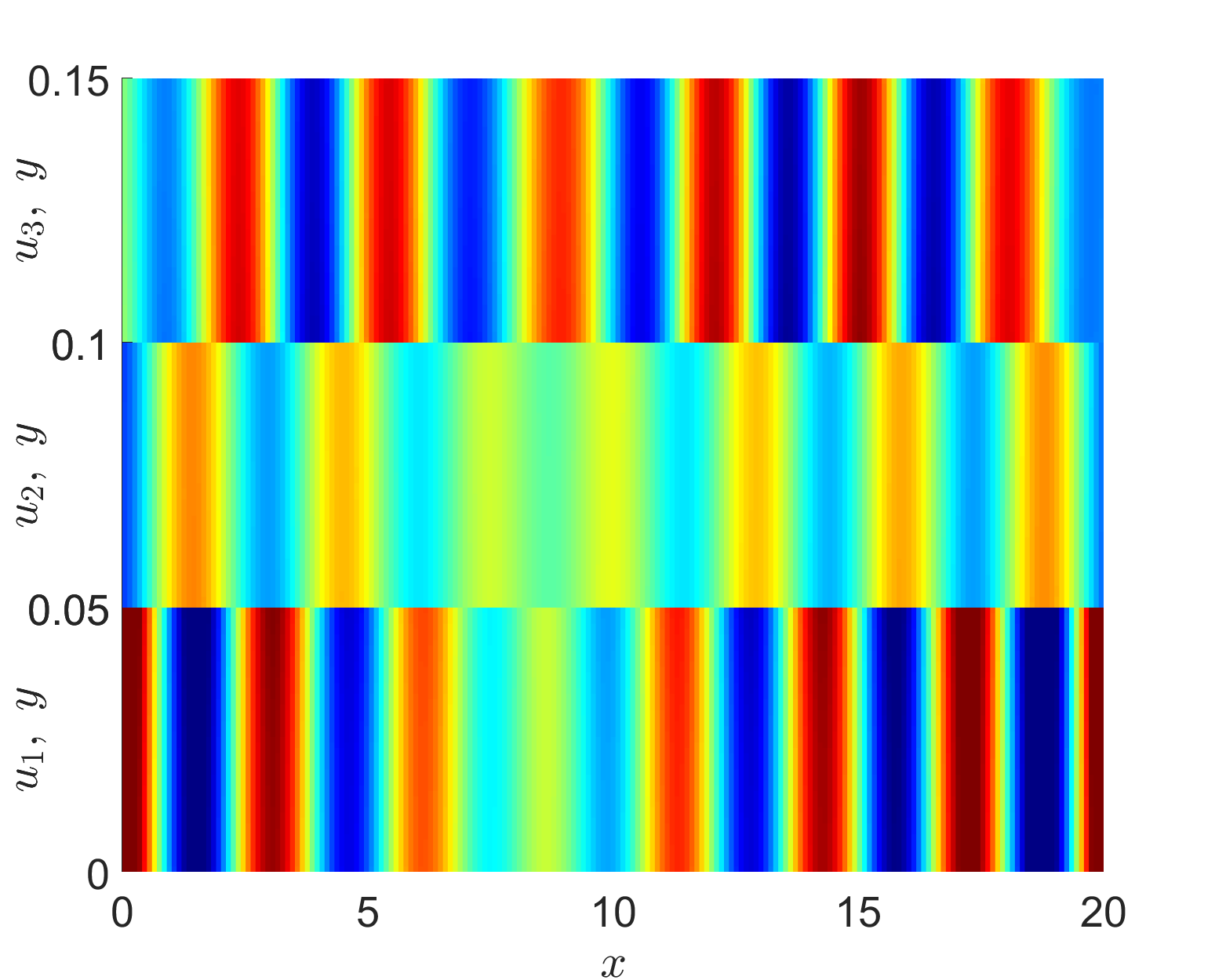}
        \caption{$t=42$}
        
    \end{subfigure}
    \begin{subfigure}[b]{0.33\textwidth}
        \includegraphics[width=\textwidth]{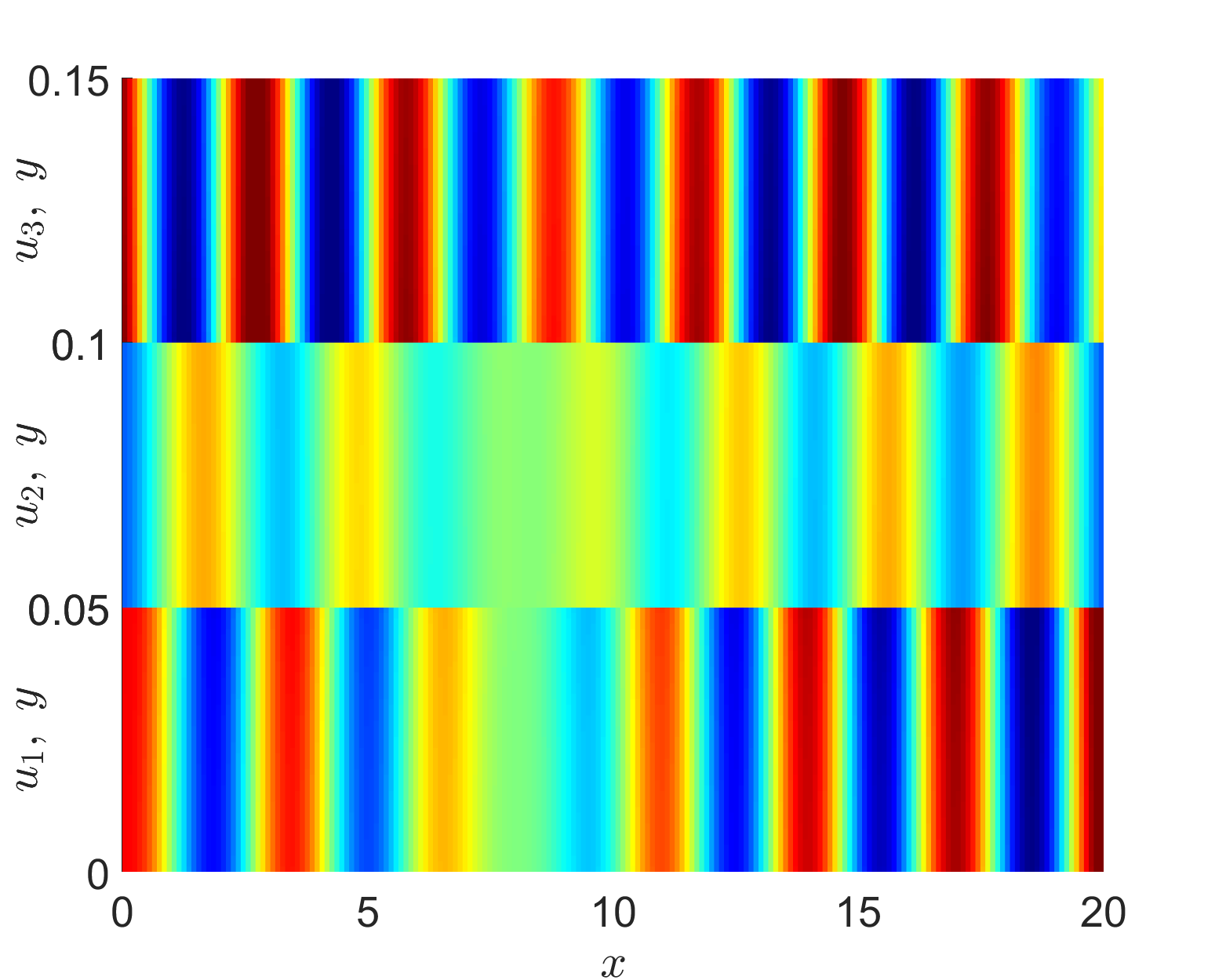}
        \caption{$t=42.5$}
        
    \end{subfigure}
    \begin{subfigure}[b]{0.33\textwidth}
        \includegraphics[width=\textwidth]{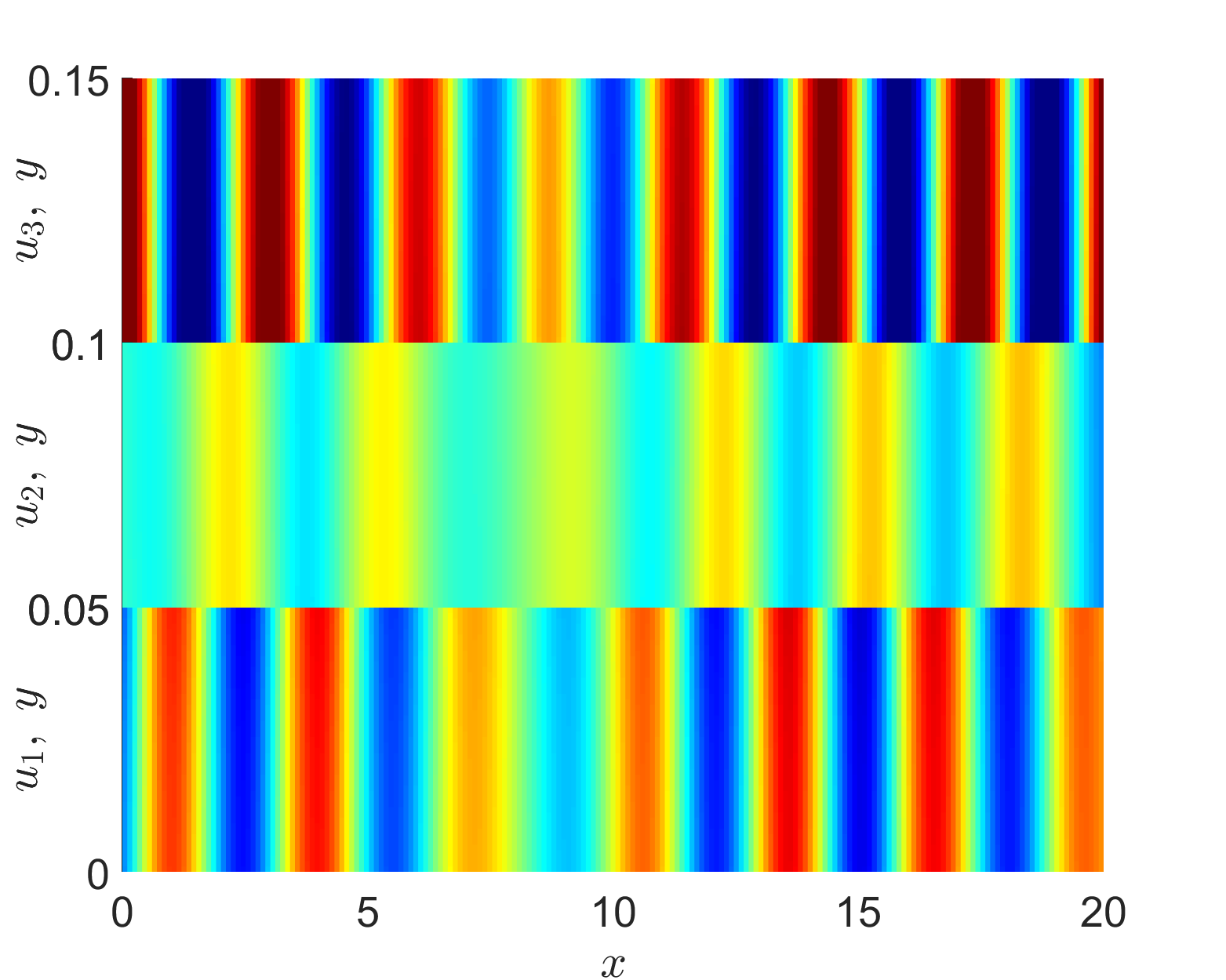}
        \caption{$t=43$}
        
    \end{subfigure}
    \begin{subfigure}[b]{0.33\textwidth}
        \includegraphics[width=\textwidth]{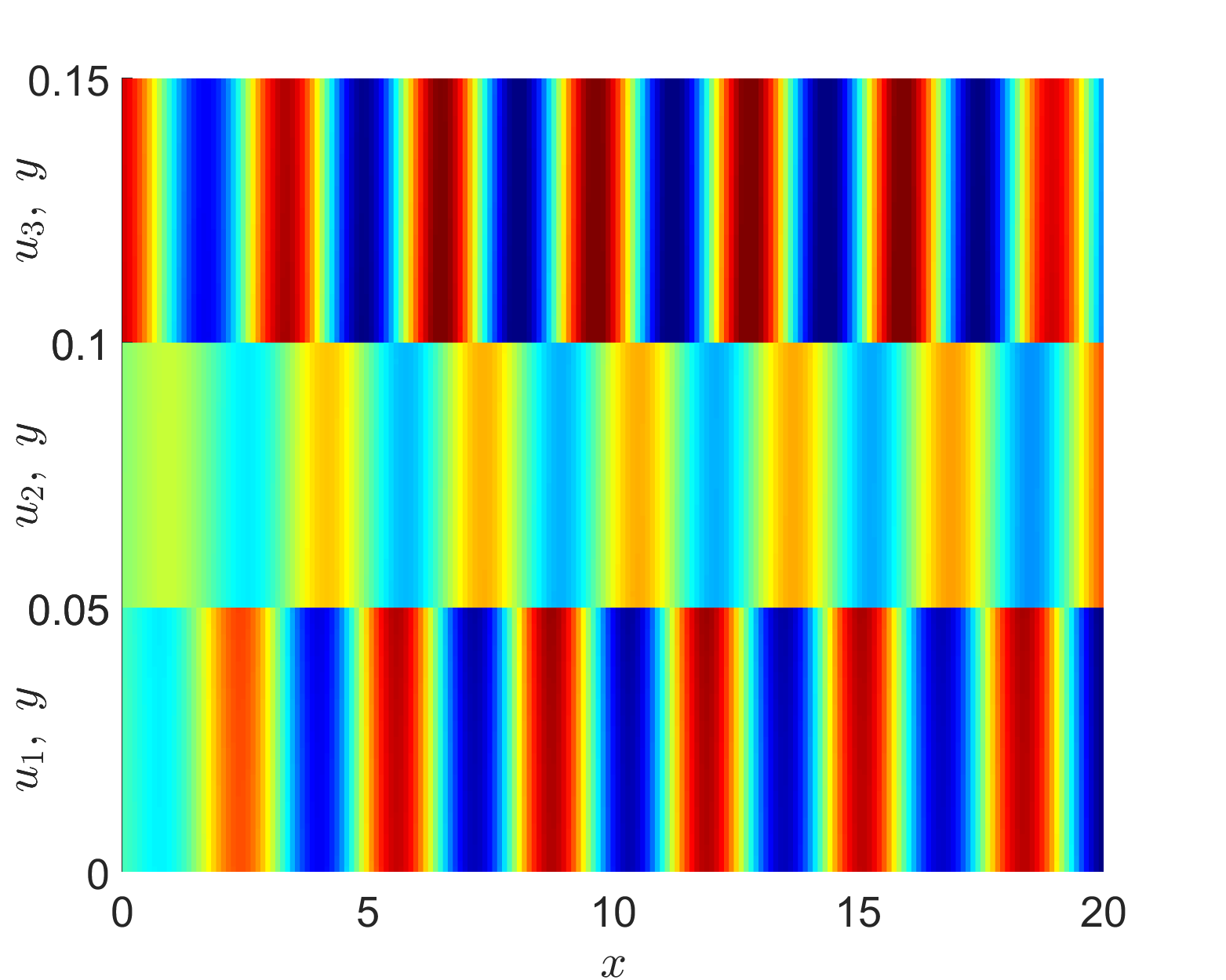}
        \caption{$t=99$}
        
    \end{subfigure}
    \begin{subfigure}[b]{0.33\textwidth}
        \includegraphics[width=\textwidth]{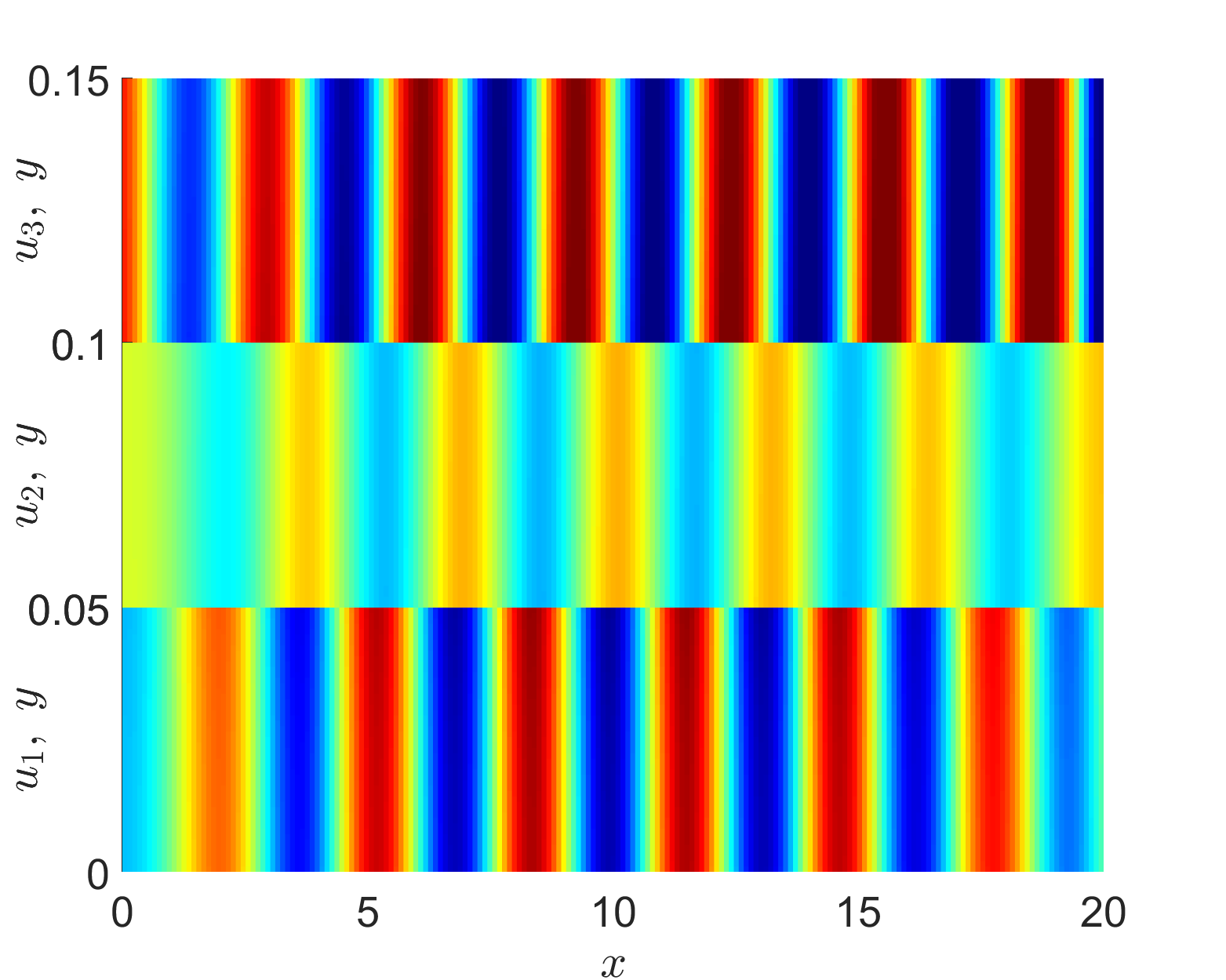}
        \caption{$t=99.5$}
        
    \end{subfigure}
    \begin{subfigure}[b]{0.33\textwidth}
        \includegraphics[width=\textwidth]{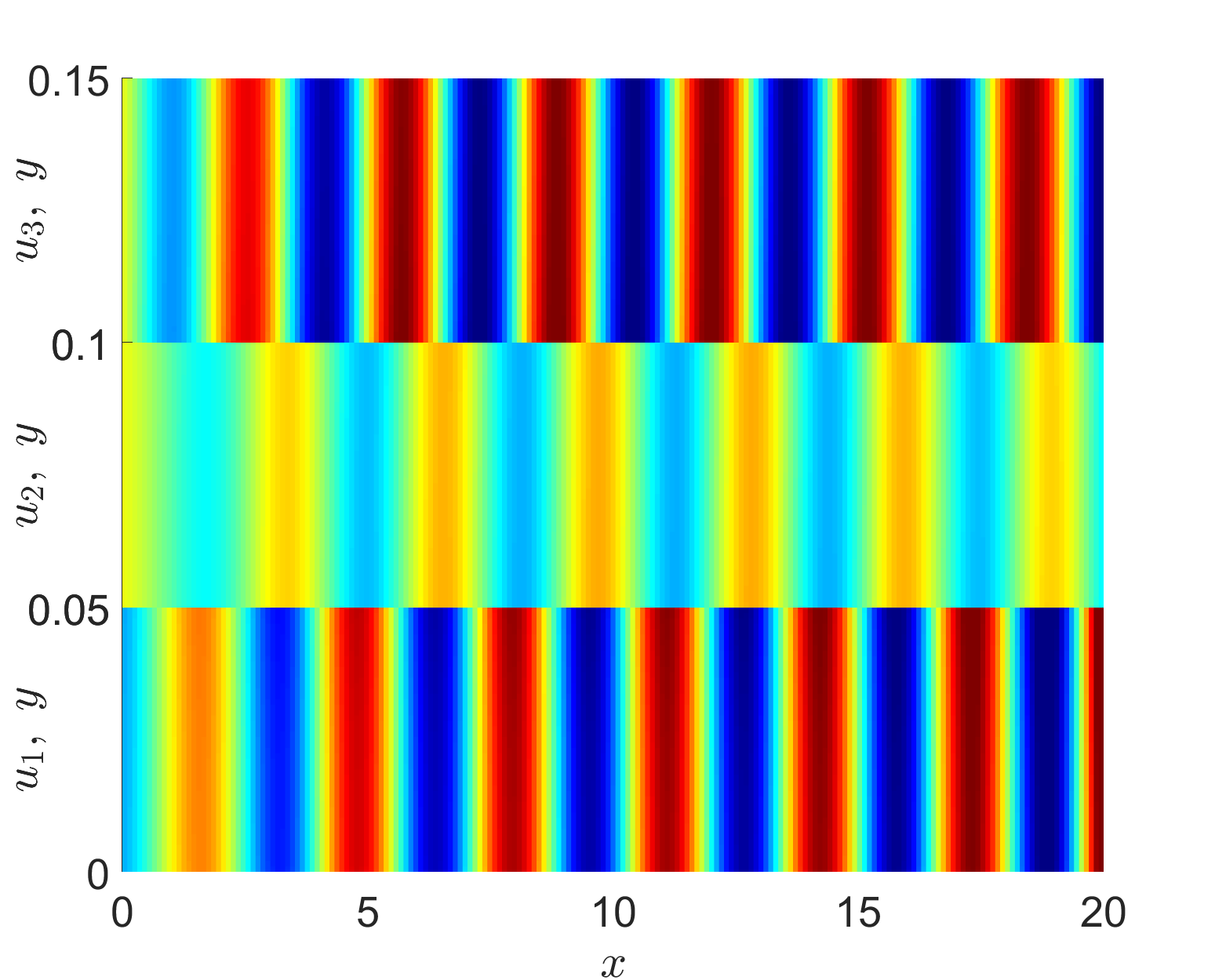}
        \caption{$t=100$}
        
    \end{subfigure}
\end{minipage}
\hspace{-0.02\textwidth}
\begin{minipage}{0.07\textwidth}
\includegraphics[width=\textwidth,height=0.71\textheight]{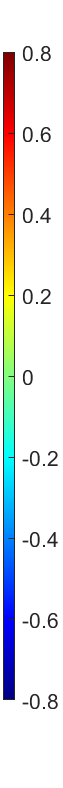}
\end{minipage}
\caption{Spatiotemporal evolution of equations~\Cref{main_multilayer_eq,neumann_boundary_conditions,coupling_boundary_conditions} for the Turing-wave model given in \Cref{system_table}. The parameters are given by $D_1 = D_3 = 0.1$, $D_2 = 1.0$, $L = 20$, $H=0.05$, $\eta=0.1$ ($\hat{\eta}=2$), with kinetic parameters $p_1 = -3$, $p_2 = -5$, $p_3 = -1$, $p_4 = 8$, $p_5 = 10$, $p_6 = 4$, and $p_7 = -2$. A continuous animation of this wave propagation is available at \citep{SMMLM2026code}.}
\label{fig:turing_wave}
\end{figure}

\begin{figure}
\begin{minipage}{0.95\textwidth}
    \centering
    \begin{subfigure}[b]{0.329\textwidth}
        \includegraphics[width=\textwidth]{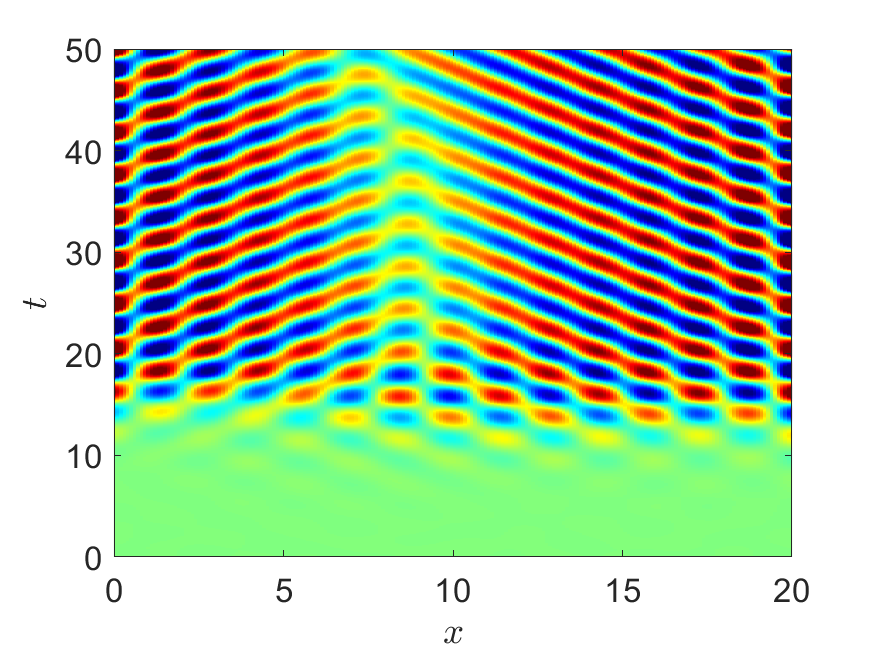}
        \caption{}
        
    \end{subfigure}
    \begin{subfigure}[b]{0.329\textwidth}
        \includegraphics[width=\textwidth]{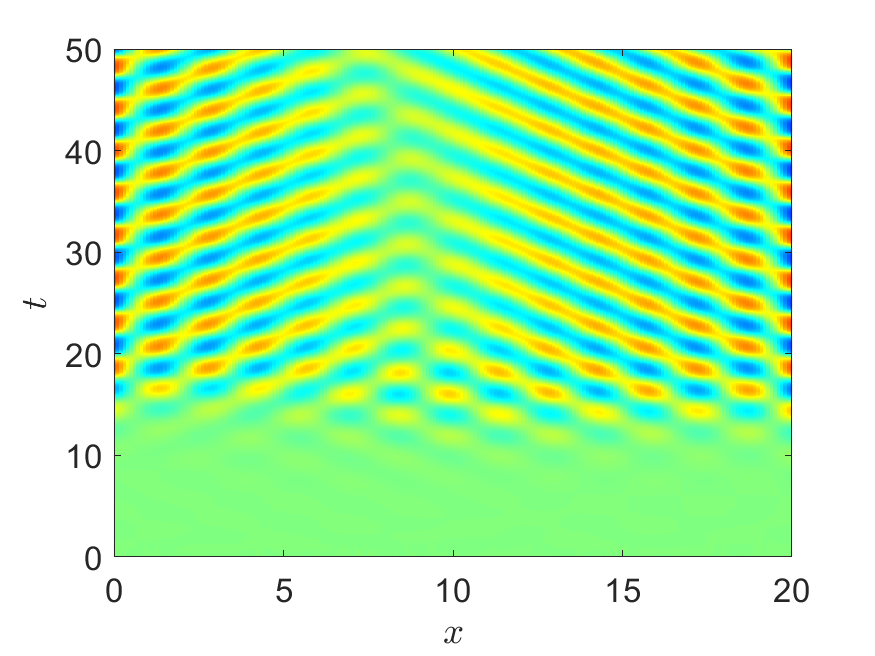}
        \caption{}
        
    \end{subfigure}
    \begin{subfigure}[b]{0.329\textwidth}
        \includegraphics[width=\textwidth]{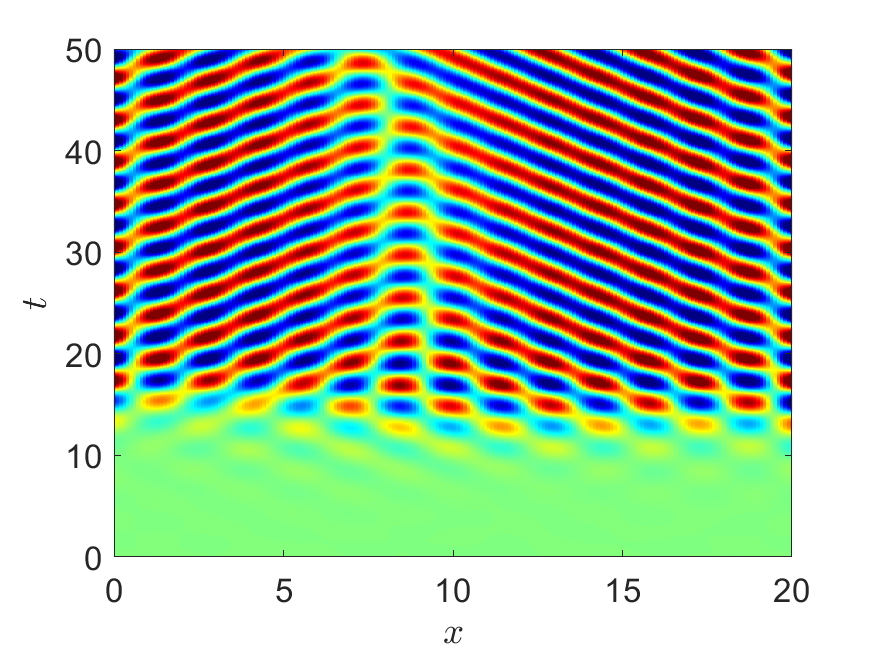}
        \caption{}
        
    \end{subfigure}
    \begin{subfigure}[b]{0.329\textwidth}
        \includegraphics[width=\textwidth]{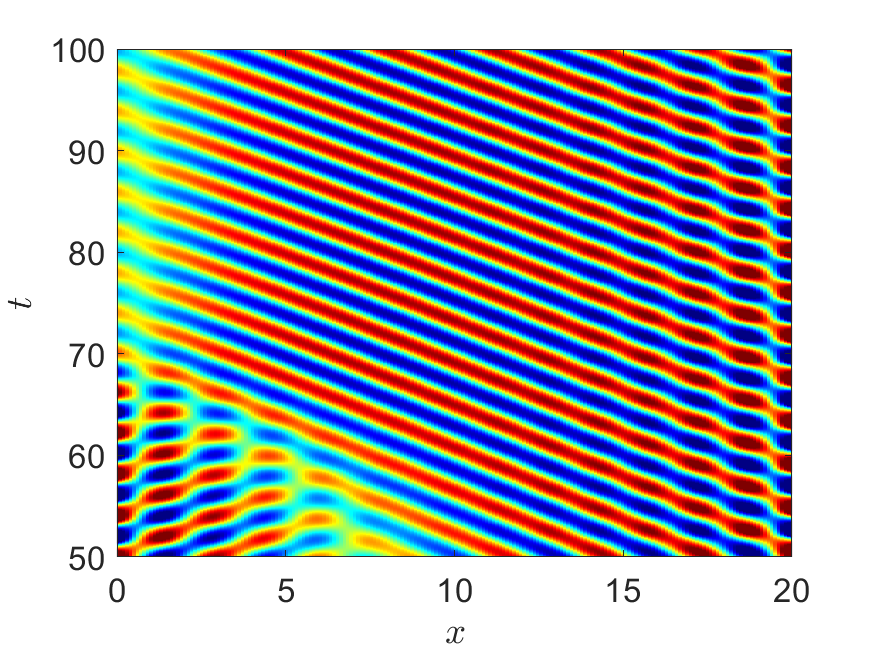}
        \caption{}
        
    \end{subfigure}
    \begin{subfigure}[b]{0.329\textwidth}
        \includegraphics[width=\textwidth]{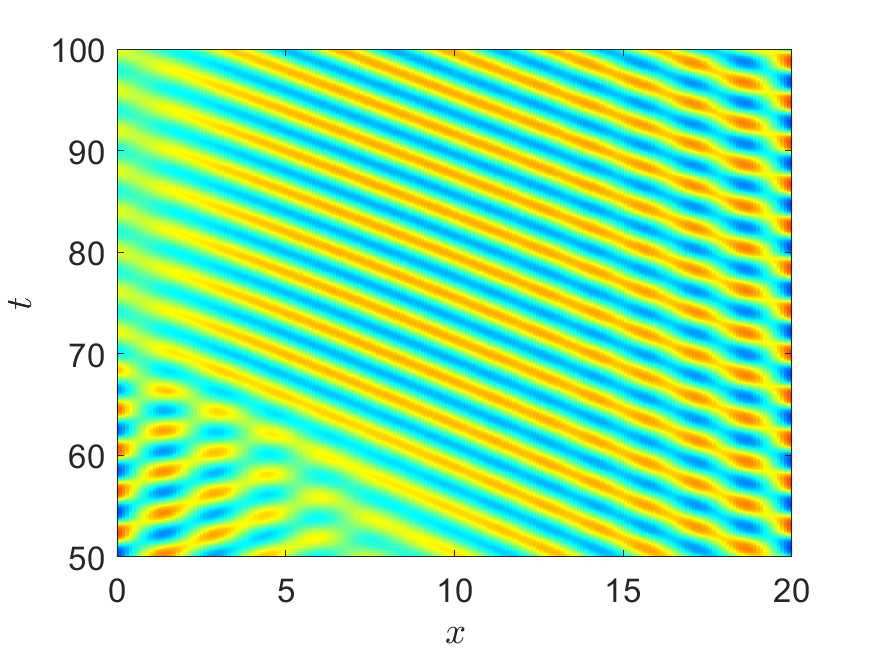}
        \caption{}
        
    \end{subfigure}
    \begin{subfigure}[b]{0.329\textwidth}
        \includegraphics[width=\textwidth]{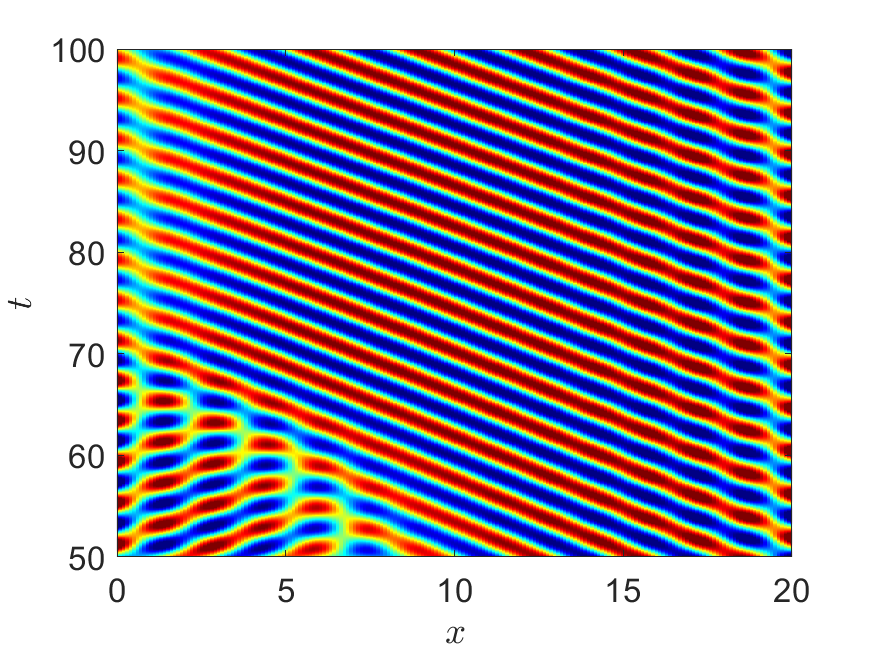}
        \caption{}
        
    \end{subfigure}
\end{minipage}
\hspace{-0.02\textwidth}
\begin{minipage}{0.04\textwidth}
\includegraphics[width=\textwidth,height=0.4\textheight]{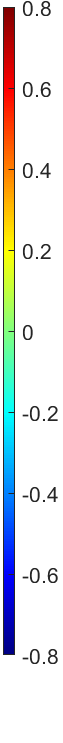}
\end{minipage}
\caption{Spatiotemporal kymographs of the simulations shown in \Cref{fig:turing_wave}. The plots show $y$-averaged concentrations, $\bar{u}_i(x,t)$ for $i=1,2,3$ (left to right). (a)--(c) correspond to early-time dynamics ($t \in [0,50]$); (d)--(f) correspond to late-time dynamics ($t \in [50,100]$).}
\label{fig:turing_wave_kymo}
\end{figure}

\begin{figure}
    \centering
    \begin{subfigure}[b]{0.45\textwidth}
        \includegraphics[width=\textwidth]{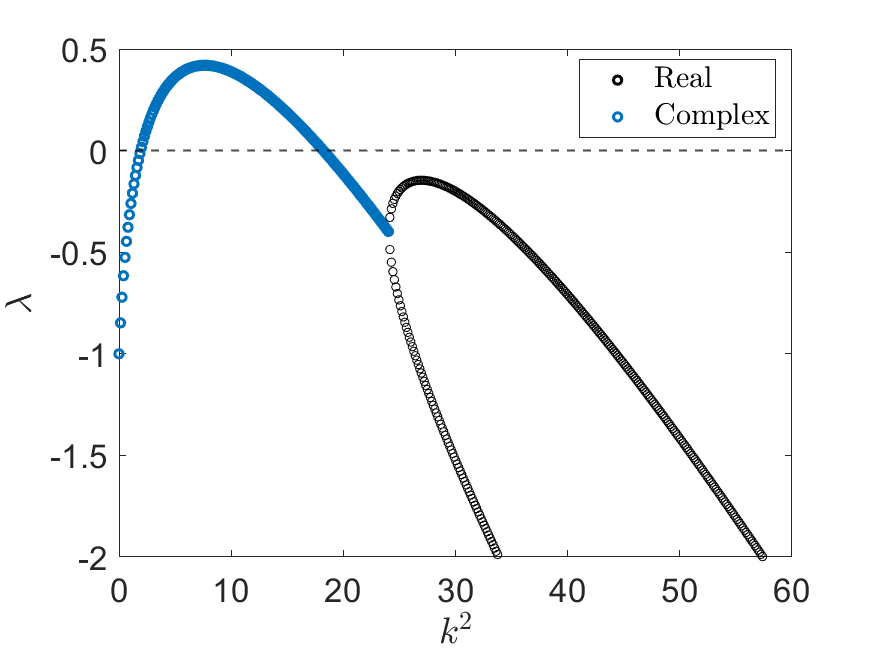}
        \caption{$\hat{\eta}=1$}
        
    \end{subfigure}
    \begin{subfigure}[b]{0.45\textwidth}
        \includegraphics[width=\textwidth]{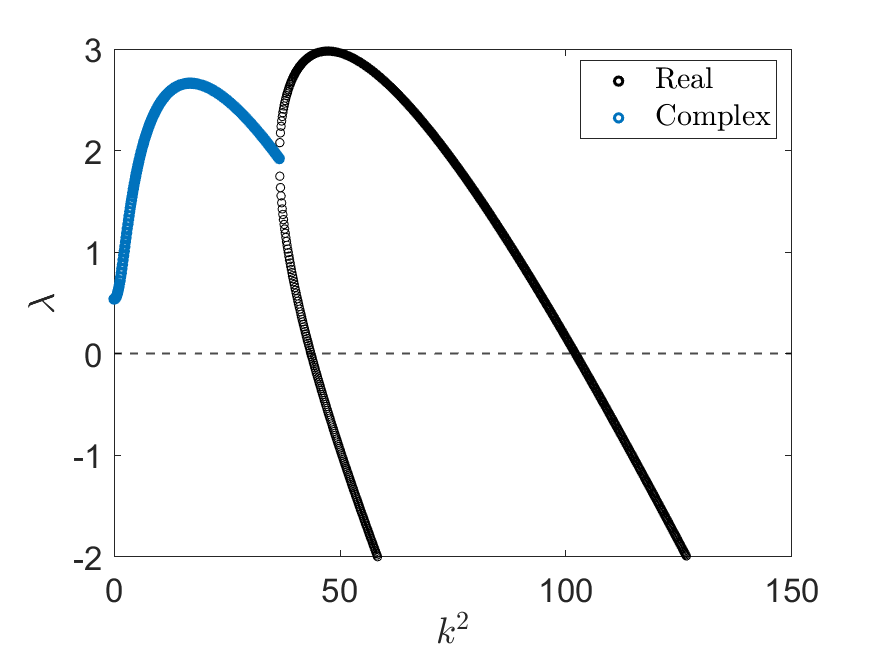}
        \caption{$\hat{\eta}=2$}
        
    \end{subfigure}
\caption{Dispersion relations for the reduced model given in \Cref{eq:reduced_wave_model}. All other parameters are as in \Cref{fig:turing_wave}.}
\label{fig:turing_wave_dispersion}
\end{figure}

The sequential snapshots in \Cref{fig:turing_wave}(a)--(c) show the evolution of the spatial pattern over time, with changes in the position of peaks suggesting lateral propagation, consistent with a coherent travelling-wave state in the full two-dimensional geometry (see also the linked video, which clearly demonstrates wave propagation). These dynamics demonstrate that nonlinear layering can generate directed spatiotemporal behaviour in a single-morphogen system without external advection.

The linear stability analysis of~\Cref{eq:reduced_wave_model} yields eigenvalues of the form $\lambda(k_q)=\alpha(k_q)\pm i\omega(k_q)$, indicating oscillatory modes. The corresponding dispersion relations (\Cref{fig:turing_wave_dispersion}) show that for $\hat{\eta}=1$, the $k_q=0$ mode remains stable while a band of nonzero wavenumbers satisfies $\Re(\lambda(k_q))>0$, consistent with a Turing--wave type instability. In contrast, for $\hat{\eta}=2$, the instability extends to $k_q=0$, and the leading eigenvalue is real and positive. Complex eigenvalues with positive real parts are also present over a range of nonzero wavenumbers, so oscillatory spatial modes remain linearly unstable.

In the full two-dimensional simulations, however, spatiotemporal wave patterns are observed only for $\hat{\eta}=2$, as illustrated in \Cref{fig:turing_wave,fig:turing_wave_kymo}. This highlights a quantitative discrepancy between the reduced and full models, where the onset of observable Turing-wave dynamics is shifted to larger effective coupling, consistent with the behaviour seen in other examples. 



\section{Conclusion}\label{Sec:Conclusion}

We have shown that a single diffusing scalar field distributed across nonlinearly coupled layers can generate spatially and temporally inhomogeneous dynamics that are impossible for scalar reaction-diffusion equations on a single homogeneous convex domain with no-flux boundaries. Starting from a general $N$-layer two-dimensional model, we derived a thin-layer reduction to an effective $N$-component reaction-diffusion system on a one-dimensional domain. This reduced model makes the origin of the dynamics transparent, and indeed is how we arrived at the exemplar systems shown here. The interfacial flux laws induce an effective coupling matrix whose structure determines whether pattern-forming instabilities can occur. In particular, purely diffusive interlayer coupling is insufficient, whereas reactive coupling can generate Turing, Hopf, and Turing-wave instabilities. These mechanisms were then illustrated through explicit examples, and the corresponding solution behaviours were shown numerically to persist in the full two-dimensional layered geometry over a substantial range of parameter values, albeit with some notable differences from the 1D reduction, even in the case of thin domains.

Our results also provide guidance on how and why the thin-layer description succeeds, and where it begins to fail. In regimes where the layers are sufficiently thin, the dominant dynamics are similar to the reduced one-dimensional system, so the layered medium behaves like an effective multi-morphogen reaction-diffusion model. As the layer thickness and coupling strength increase, however, vertical gradients (perpendicular to the interacting boundary) become significant and the full two-dimensional geometry admits behaviours that are not well captured by the asymptotic reduction. In this sense, we identify both a reduced instability mechanism and the geometric limits of its validity. 

A more fundamental limitation of this reduction is that nonlinear fluxes $g_i$ will not in general lead to continuous concentrations across the interfaces, invalidating the use of \Cref{limit_result} in deriving the reduced model even when asymptotically small layer thickness and coupling are used. Resolving this requires a much more detailed analysis of boundary layers, and it is not obvious if it can be carried out in the same generality as the crude reduction done here. This incompatibility between the 1D and 2D models is also apparent in the lack of a spatially homogeneous equilibrium in the 2D form of the autocatalytic model, leading to the system only admitting spatially-variable equilibria even far away from any pattern-forming parameter regime. Such heterogeneous steady states are then `non-equilibrium' in a thermodynamic sense (as the fluxes remain nonzero), indicating that the nonlinear coupling used here inherently drives the system towards some kind of dissipative structure \cite{prigogine1978time}.

More broadly, the main message is multiscale. The model begins with a single morphogen at the local layer scale, but the combination of geometry, diffusion, and nonlinear interfacial exchange produces an effective larger-scale system with the dynamical freedom usually associated with multiple interacting species. The complexity therefore does not come from adding extra chemical species, but from resolving how information is transmitted across interfaces and then asking what coarse-grained dynamics this transmission induces. From this viewpoint, the interface laws are not auxiliary boundary conditions. They are constitutive ingredients that determine how microscale heterogeneity is converted into emergent macroscale behaviour.

This perspective extends well beyond the specific examples considered here. Many biological and physical systems are inherently layered, compartmentalised, or organised through coupling between active and passive regions, and in such settings the admissible dynamics are shaped not only by the local kinetics, but also by how transport and exchange occur across interfaces. Representative examples include developmental systems with epithelial-mesenchymal or epidermal-dermal coupling, bacterial colonies or synthetic signalling layers grown on agar where a chemically active upper layer exchanges diffusible species with a comparatively passive substrate, and membrane-cytosol polarity models in which reactions are localised to an active membrane while diffusion takes place throughout the surrounding bulk \citep{stratified,gomez_pattern_2021,PaquinLefebvre2020}. In each case, the underlying transport problem is not that of a single homogeneous medium, but of a spatially structured system in which geometry and interfacial laws help determine which instabilities are possible, how signals propagate between regions, and what effective macroscale dynamics emerge.

The present framework provides a minimal demonstration that layering can itself act as a mechanism for pattern formation and spatiotemporal complexity. More generally, it suggests that spatial organisation across scales may play a role analogous to the introduction of additional reacting species, by creating effective dynamical degrees of freedom through interfacial exchange. This opens several natural directions for future work, including the rigorous analysis of well-posedness and long-time behaviour in evolving bulk-surface settings \citep{Caetano2025}, the derivation of effective interfacial coupling laws from finer-scale continuum and thermodynamic descriptions \citep{duda2023modelling}, and extensions to more general geometries where symmetry, curvature, and bulk-surface coupling influence both onset and selection of patterned states \citep{VillarSepulveda2026,Woolley2025Polygons}. Across all of these settings, the broader lesson is the same: multiscale spatial structure is not merely a geometric complication, but can be a generative dynamical ingredient in its own right.

\bibliographystyle{unsrtnat}
\bibliography{refs.bib}

@article{stratified,
  title={Turing patterning in stratified domains},
  author={Krause, Andrew L and Klika, V{\'a}clav and Halatek, Jacob and Grant, Paul K and Woolley, Thomas E and Dalchau, Neil and Gaffney, Eamonn A},
  journal={Bulletin of {M}athematical {B}iology},
  volume={82},
  number={10},
  pages={136},
  year={2020},
  publisher={Springer}
}

@article{matano1979asymptotic,
  title={Asymptotic behavior and stability of solutions of semilinear diffusion equations},
  author={Matano, Hiroshi},
  journal={Publications of the Research Institute for Mathematical Sciences},
  volume={15},
  number={2},
  pages={401--454},
  year={1979},
  publisher={Research Institute forMathematical Sciences}
}

@article{casten1978instability,
  title={Instability results for reaction diffusion equations with Neumann boundary conditions},
  author={Casten, Richard G and Holland, Charles J},
  journal={Journal of Differential Equations},
  volume={27},
  number={2},
  pages={266--273},
  year={1978},
  publisher={Elsevier}
}

@article{diez2024turing,
  title={Turing pattern formation in reaction-cross-diffusion systems with a bilayer geometry},
  author={Diez, Antoine and Krause, Andrew L and Maini, Philip K and Gaffney, Eamonn A and Seirin-Lee, Sungrim},
  journal={Bulletin of {M}athematical {B}iology},
  volume={86},
  number={2},
  pages={13},
  year={2024},
  publisher={Springer}
}

@article{satnoianu2000turing,
  title={Turing instabilities in general systems},
  author={Satnoianu, Razvan A and Menzinger, Michael and Maini, Philip K},
  journal={Journal of {M}athematical {B}iology},
  volume={41},
  number={6},
  pages={493--512},
  year={2000},
  publisher={Springer}
}

@article{villar2025designing,
  title={Designing reaction-cross-diffusion systems with {T}uring and wave instabilities},
  author={Villar-Sep{\'u}lveda, Edgardo and Champneys, Alan R and Krause, Andrew L},
  journal={Journal of {M}athematical {B}iology},
  volume={91},
  number={4},
  pages={37},
  year={2025},
  publisher={Springer}
}

@Article{Schnakenberg-1979-SCR,
  Title                    = {Simple chemical reaction systems with limit cycle behaviour.},
  Author                   = {Schnakenberg, J.},
  Journal                  = {J. Theor. Biol.},
  Year                     = {1979},
  Number                   = {3},
  Pages                    = {389--400},
  Volume                   = {81}
}

@article{Servedio2014,
  author  = {Servedio, Maria R. and Brandvain, Yaniv and Dhole, Sumit and Fitzpatrick, Courtney L. and Goldberg, Emma E. and Stern, Caitlin A. and Van Cleve, Jeremy and Yeh, D. Justin},
  title   = {Not Just a Theory---The Utility of Mathematical Models in Evolutionary Biology},
  journal = {PLoS Biology},
  volume  = {12},
  number  = {12},
  pages   = {e1002017},
  year    = {2014},
  doi     = {10.1371/journal.pbio.1002017}
}

@article{Maini2012,
  author  = {Maini, Philip K. and Woolley, Thomas E. and Baker, Ruth E. and Gaffney, Eamonn A. and Lee, S. Seirin},
  title   = {Turing's model for biological pattern formation and the robustness problem},
  journal = {Interface Focus},
  volume  = {2},
  number  = {4},
  pages   = {487--496},
  year    = {2012},
  doi     = {10.1098/rsfs.2011.0113}
}

@article{PaquinLefebvre2020,
  author  = {Paquin-Lefebvre, Fr{\'e}d{\'e}ric and Xu, Bin and DiPietro, Kelsey L. and Lindsay, Alan E. and Jilkine, Alexandra},
  title   = {Pattern formation in a coupled membrane-bulk reaction-diffusion model for intracellular polarization and oscillations},
  journal = {Journal of Theoretical Biology},
  volume  = {497},
  pages   = {110242},
  year    = {2020},
  doi     = {10.1016/j.jtbi.2020.110242}
}

@article{Caetano2025,
  author  = {Caetano, Diogo and Elliott, Charles M. and Tang, Bao Quoc},
  title   = {Bulk-surface systems on evolving domains},
  journal = {Journal of Evolution Equations},
  volume  = {25},
  pages   = {103},
  year    = {2025},
  doi     = {10.1007/s00028-025-01130-5}
}

@article{VillarSepulveda2026,
  author  = {Villar-Sep{\'u}lveda, Edgardo and Champneys, Alan R. and Cusseddu, Davide and Madzvamuse, Anotida},
  title   = {Pattern formation of bulk-surface reaction-diffusion systems in a ball},
  journal = {SIAM Journal on Applied Mathematics},
  volume  = {86},
  number  = {1},
  pages   = {21--51},
  year    = {2026},
  doi     = {10.1137/24M1671037}
}

@article{villar2023general,
  title={General conditions for {T}uring and wave instabilities in reaction-diffusion systems},
  author={Villar-Sep{\'u}lveda, Edgardo and Champneys, Alan R},
  journal={Journal of {M}athematical {B}iology},
  volume={86},
  number={3},
  pages={39},
  year={2023},
  publisher={Springer}
}

@article{catlla2012instabilities,
  title={Instabilities and patterns in coupled reaction-diffusion layers},
  author={Catll{\'a}, Anne J and McNamara, Amelia and Topaz, Chad M},
  journal={Physical Review E—Statistical, Nonlinear, and Soft Matter Physics},
  volume={85},
  number={2},
  pages={026215},
  year={2012},
  publisher={APS}
}

@article{satnoianu2005some,
  title={Some remarks on matrix stability with application to {T}uring instability},
  author={Satnoianu, Razvan A and  van den Driessche, Pauline},
  journal={Linear algebra and its applications},
  volume={398},
  pages={69--74},
  year={2005},
  publisher={Elsevier}
}

@article{neubert2002transient,
  title={Transient dynamics and pattern formation: reactivity is necessary for {T}uring instabilities},
  author={Neubert, Michael G and Caswell, Hal and Murray, JD},
  journal={Mathematical {B}iosciences},
  volume={175},
  number={1},
  pages={1--11},
  year={2002},
  publisher={Elsevier}
}

@article{porter2016dynamical,
  title={Dynamical systems on networks},
  author={Porter, Mason A and Gleeson, James P},
  journal={Frontiers in Applied Dynamical Systems: Reviews and Tutorials},
  volume={4},
  pages={29},
  year={2016},
  publisher={Springer}
}

@book{thomee2007galerkin,
  title={Galerkin finite element methods for parabolic problems},
  author={Thom{\'e}e, Vidar},
  volume={25},
  year={2007},
  publisher={Springer Science \& Business Media}
}

@book{hundsdorfer2013numerical,
  title={Numerical solution of time-dependent advection-diffusion-reaction equations},
  author={Hundsdorfer, Willem and Verwer, Jan G},
  volume={33},
  year={2013},
  publisher={Springer Science \& Business Media}
}

@article{gaffney_spatial_2023,
  title = {Spatial Heterogeneity Localizes {{Turing}} Patterns in Reaction-Cross-Diffusion Systems},
  author = {Gaffney, Eamonn A. and Krause, Andrew L. and Maini, Philip K. and Wang, Chenyuan and {Mathematical Sciences Department, Durham University, Upper Mountjoy Campus, Stockton Rd, Durham DH1 3LE, United Kingdom}},
  date = {2023},
  journal = {Discrete Contin. Dyn. Syst. Ser. B},
  volume = {In press},
  issn = {1531-3492, 1553-524X},
  doi = {10.3934/dcdsb.2023053},
  url = {https://www.aimsciences.org//article/doi/10.3934/dcdsb.2023053},
  urldate = {2023-05-12}
}

@article{gierer_theory_1972,
  title = {A Theory of Biological Pattern Formation},
  author = {Gierer, A. and Meinhardt, H.},
  date = {1972-12},
  Year                     = {1972},
  journal = {Kybernetik},
  shortjournal = {Kybernetik},
  volume = {12},
  number = {1},
  pages = {30--39},
  issn = {0023-5946},
  doi = {10.1007/BF00289234},
  url = {https://link.springer.com/10.1007/BF00289234},
  urldate = {2023-03-03}
}

@Book{Comsol-2021,
  author    = {COMSOL Multiphysics},
  date      = {2021},
  Year                     = {2021},
  title     = {v. 5.1},
  location  = {COMSOL AB,Stockholm, Sweden},
  publisher = {www.comsol.com},
}

@article{gomez_pattern_2021,
  title = {Pattern Forming Systems Coupling Linear Bulk Diffusion to Dynamically Active Membranes or Cells},
  author = {Gomez, D. and Iyaniwura, S. and Paquin-Lefebvre, F. and Ward, M. J.},
  date = {2021-12-27},
  Year                     = {2021},
  journal = {Phil. Trans. R. Soc. A.},
  shortjournal = {Phil. Trans. R. Soc. A.},
  volume = {379},
  number = {2213},
  pages = {20200276},
  issn = {1364-503X, 1471-2962},
  doi = {10.1098/rsta.2020.0276},
  url = {https://royalsocietypublishing.org/doi/10.1098/rsta.2020.0276},
  urldate = {2023-02-28}
}

@article{duda2023modelling,
  title={Modelling of surface reactions and diffusion mediated by bulk diffusion},
  author={Duda, Fernando P and Forte Neto, Francisco S and Fried, Eliot},
  journal={Philosophical Transactions of the Royal Society A: Mathematical, Physical and Engineering Sciences},
  volume={381},
  number={2263},
  year={2023},
  publisher={The Royal Society}
}

@article{Krause2020WKBJ,
  author  = {Krause, Andrew L. and Klika, V{\'a}clav and Woolley, Thomas E. and Gaffney, Eamonn A.},
  title   = {From one pattern into another: Analysis of Turing patterns in heterogeneous domains via WKBJ},
  journal = {Journal of the Royal Society Interface},
  volume  = {17},
  number  = {162},
  pages   = {20190621},
  year    = {2020},
  doi     = {10.1098/rsif.2019.0621}
}

@article{Woolley2025Polygons,
  author  = {Woolley, Thomas E.},
  title   = {Pattern formation on regular polygons and circles},
  journal = {Journal of Nonlinear Science},
  volume  = {35},
  pages   = {5},
  year    = {2025},
  doi     = {10.1007/s00332-024-10096-6}
}

@article{TuringHopfSchnakenberg,
title = {The amplitude system for a Simultaneous short-wave Turing and long-wave Hopf instability},
journal = {Discrete and Continuous Dynamical Systems - S},
volume = {15},
number = {9},
pages = {2657-2672},
year = {2022},
issn = {1937-1632},
doi = {10.3934/dcdss.2021119},
url = {https://www.aimsciences.org/article/id/17fad18c-9de1-4651-9904-d4998bbf49a4},
author = {Guido Schneider and Matthias Winter}
}

@article{novak2008design,
  title={Design principles of biochemical oscillators},
  author={Nov{\'a}k, B{\'e}la and Tyson, John J},
  journal={Nature reviews Molecular cell biology},
  volume={9},
  number={12},
  pages={981--991},
  year={2008},
  publisher={Nature Publishing Group UK London}
}

@article{BronsardKohn,
author = {Bronsard, Lia and Kohn, Robert V.},
title = {On the slowness of phase boundary motion in one space dimension},
journal = {Communications on Pure and Applied Mathematics},
volume = {43},
number = {8},
pages = {983-997},
year = {1990}
}

@article{walker2023visualpde,
  title={VisualPDE: rapid interactive simulations of partial differential equations},
  author={Walker, Benjamin J and Townsend, Adam K and Chudasama, Alexander K and Krause, Andrew L},
  journal={Bulletin of Mathematical Biology},
  volume={85},
  number={11},
  pages={113},
  year={2023},
  publisher={Springer}
}

@article{levine_membranebound_2005,
  title = {Membrane-Bound {{Turing}} Patterns},
  author = {Levine, Herbert and Rappel, Wouter-Jan},
  date = {2005-12-19},
  journal = {Phys. Rev. E},
  shortjournal = {Phys. Rev. E},
  volume = {72},
  number = {6},
  pages = {061912},
  issn = {1539-3755, 1550-2376},
  doi = {10.1103/PhysRevE.72.061912},
  url = {https://link.aps.org/doi/10.1103/PhysRevE.72.061912},
  urldate = {2023-02-28}
}

@article{madzvamuse_stability_2015,
  title = {Stability Analysis and Simulations of Coupled Bulk-Surface Reaction–Diffusion Systems},
  author = {Madzvamuse, Anotida and Chung, Andy H. W. and Venkataraman, Chandrasekhar},
  date = {2015-03},
  journal = {Proc. R. Soc. A.},
  volume = {471},
  number = {2175},
  pages = {20140546},
  issn = {1364-5021, 1471-2946},
  doi = {10.1098/rspa.2014.0546},
  url = {https://royalsocietypublishing.org/doi/10.1098/rspa.2014.0546},
  urldate = {2023-03-02}
}

@book{murray_mathematical_2003,
  title = {Mathematical {{Biology}}: {{II}}: {{Spatial Models}} and {{Biomedical Applications}}},
  shorttitle = {Mathematical {{Biology}}},
  author = {Murray, J. D.},
  date = {2003},
  Year       = {2003},
  series = {Interdisciplinary {{Applied Mathematics}}},
  edition = {3},
  volume = {18},
  publisher = {{Springer New York}},
  location = {{New York, NY}},
  doi = {10.1007/b98869},
  url = {http://link.springer.com/10.1007/b98869},
  urldate = {2023-02-24}
}

@article{paquin-lefebvre_pattern_2019,
  title = {Pattern {{Formation}} and {{Oscillatory Dynamics}} in a {{Two-Dimensional Coupled Bulk-Surface Reaction-Diffusion System}}},
  author = {Paquin-Lefebvre, Frédéric and Nagata, Wayne and Ward, Michael J.},
  date = {2019},
  journal = {SIAM J. Appl. Dyn. Syst.},
  volume = {18},
  number = {3},
  pages = {1334--1390},
  doi = {10.1137/18M1213737}
}

@article{paquin-lefebvre_pattern_2020,
  title = {Pattern Formation in a Coupled Membrane-Bulk Reaction-Diffusion Model for Intracellular Polarization and Oscillations},
  author = {Paquin-Lefebvre, Frédéric and Xu, Bin and DiPietro, Kelsey L. and Lindsay, Alan E. and Jilkine, Alexandra},
  date = {2020-07},
  journal = {J. Theor. Biol.},
  volume = {497},
  pages = {110242},
  doi = {10.1016/j.jtbi.2020.110242}
}

@article{ratz_symmetry_2014,
  title = {Symmetry Breaking in a Bulk–Surface Reaction–Diffusion Model for Signalling Networks},
  author = {Rätz, Andreas and Röger, Matthias},
  date = {2014-08-01},
  journal = {Nonlinearity},
  shortjournal = {Nonlinearity},
  volume = {27},
  number = {8},
  pages = {1805--1827},
  doi = {10.1088/0951-7715/27/8/1805},
  url = {https://iopscience.iop.org/article/10.1088/0951-7715/27/8/1805},
  urldate = {2023-02-28}
}

@article{ratz_turingtype_2015,
  title = {Turing-Type Instabilities in Bulk–Surface Reaction–Diffusion Systems},
  author = {Rätz, Andreas},
  date = {2015-12},
  journal = {Journal of Computational and Applied Mathematics},
  shortjournal = {Journal of Computational and Applied Mathematics},
  volume = {289},
  pages = {142--152},
  issn = {03770427},
  doi = {10.1016/j.cam.2015.02.050},
  url = {https://linkinghub.elsevier.com/retrieve/pii/S0377042715001272},
  urldate = {2023-02-28}
}

@article{turing_chemical_1952,
  title = {The Chemical Basis of Morphogenesis},
  author = {Turing, Alan},
  date = {1952-08-14},
  Year = {1979},
  journal = {Phil. Trans. R. Soc. Lond. B},
  shortjournal = {Phil. Trans. R. Soc. Lond. B},
  volume = {237},
  number = {641},
  pages = {37--72},
  issn = {2054-0280},
  doi = {10.1098/rstb.1952.0012},
  url = {https://royalsocietypublishing.org/doi/10.1098/rstb.1952.0012},
  urldate = {2023-03-02}
}

@book{horn2012matrix,
  title={Matrix analysis},
  author={Horn, Roger A and Johnson, Charles R},
  year={2012},
  publisher={Cambridge university press}
}

@article{neubert1997alternatives,
  title={Alternatives to resilience for measuring the responses of ecological systems to perturbations},
  author={Neubert, Michael G and Caswell, Hal},
  journal={Ecology},
  volume={78},
  number={3},
  pages={653--665},
  year={1997},
  publisher={Wiley Online Library}
}

@misc{SMMLM2026code,
howpublished={\url{https://github.com/MahashriN/Single_Morphogen_Multi_Layered_Model}}
}

@article{weevers2025mechanochemical,
  title={Mechanochemical patterning localizes the organizer of a luminal epithelium},
  author={Weevers, Sera L and Falconer, Alistair D and Mercker, Moritz and Sadeghi, Hajar and Rozema, David and Ferenc, Jaroslav and Ma{\^\i}tre, Jean-Leon and Ott, Albrecht and Oelz, Dietmar B and Marciniak-Czochra, Anna and others},
  journal={Science Advances},
  volume={11},
  number={26},
  pages={eadu2286},
  year={2025},
  publisher={American Association for the Advancement of Science}
}

@article{veerman2021beyond,
  title={Beyond Turing: far-from-equilibrium patterns and mechano-chemical feedback},
  author={Veerman, Frits and Mercker, Moritz and Marciniak-Czochra, Anna},
  journal={Philosophical Transactions of the Royal Society A: Mathematical, Physical and Engineering Sciences},
  volume={379},
  number={2213},
  year={2021},
  publisher={The Royal Society}
}

@article{nesenberend2025curvature,
  title={Curvature induced patterns: A geometric, analytical approach to understanding a mechanochemical model},
  author={Nesenberend, Daphne and Doelman, Arjen and Veerman, Frits},
  journal={bioRxiv},
  pages={2025--04},
  year={2025},
  publisher={Cold Spring Harbor Laboratory}
}

@article{mercker2013mechanochemical,
  title={A mechanochemical model for embryonic pattern formation: coupling tissue mechanics and morphogen expression},
  author={Mercker, Moritz and Hartmann, Dirk and Marciniak-Czochra, Anna},
  journal={PloS one},
  volume={8},
  number={12},
  pages={e82617},
  year={2013},
  publisher={Public Library of Science San Francisco, USA}
}

@article{boeynaems2018protein,
  title={Protein phase separation: a new phase in cell biology},
  author={Boeynaems, Steven and Alberti, Simon and Fawzi, Nicolas L and Mittag, Tanja and Polymenidou, Magdalini and Rousseau, Frederic and Schymkowitz, Joost and Shorter, James and Wolozin, Benjamin and Van Den Bosch, Ludo and others},
  journal={Trends in cell biology},
  volume={28},
  number={6},
  pages={420--435},
  year={2018},
  publisher={Elsevier}
}

@article{kowall2025nonlinear,
  title={Nonlinear stability results for stationary solutions of reaction-diffusion-ODE systems},
  author={Kowall, Chris and Marciniak-Czochra, Anna and M{\"u}nnich, Finn},
  journal={Journal of Differential Equations},
  volume={448},
  pages={113704},
  year={2025},
  publisher={Elsevier}
}

@article{klika2012influence,
  title={The influence of receptor-mediated interactions on reaction-diffusion mechanisms of cellular self-organisation},
  author={Klika, V{\'a}clav and Baker, Ruth E and Headon, Denis and Gaffney, Eamonn A},
  journal={Bulletin of mathematical biology},
  volume={74},
  number={4},
  pages={935--957},
  year={2012},
  publisher={Springer}
}

@article{korvasova2015investigating,
  title={Investigating the Turing conditions for diffusion-driven instability in the presence of a binding immobile substrate},
  author={Korvasov{\'a}, K and Gaffney, EA and Maini, PK and Ferreira, MA and Klika, V},
  journal={Journal of theoretical biology},
  volume={367},
  pages={286--295},
  year={2015},
  publisher={Elsevier}
}

@article{menou2023physical,
  title={Physical interactions in non-ideal fluids promote Turing patterns},
  author={Menou, Lucas and Luo, Chengjie and Zwicker, David},
  journal={Journal of the Royal Society Interface},
  volume={20},
  number={204},
  year={2023},
  publisher={The Royal Society}
}

@article{krause2025pattern,
  title={Pattern Localization in the Swift--Hohenberg Equation via Slowly Varying Spatial Heterogeneity},
  author={Krause, Andrew L and Klika, V{\'a}clav and Villar-Sep{\'u}lveda, Edgardo and Champneys, Alan R and Gaffney, Eamonn A},
  journal={SIAM Journal on Applied Dynamical Systems},
  volume={24},
  number={4},
  pages={2804--2847},
  year={2025},
  publisher={SIAM}
}

@article{patterson2023spatial,
  title={Spatial dynamics with heterogeneity},
  author={Patterson, Denis D and Staver, A Carla and Levin, Simon A and Touboul, Jonathan D},
  journal={SIAM journal on applied mathematics},
  pages={S225--S248},
  year={2023},
  publisher={SIAM}
}

@article{krause2018heterogeneity,
  title={Heterogeneity induces spatiotemporal oscillations in reaction-diffusion systems},
  author={Krause, Andrew L and Klika, V{\'a}clav and Woolley, Thomas E and Gaffney, Eamonn A},
  journal={Physical Review E},
  volume={97},
  number={5},
  pages={052206},
  year={2018},
  publisher={APS}
}

@article{woolley2021bespoke,
  title={Bespoke {T}uring systems},
  author={Woolley, Thomas E and Krause, Andrew L and Gaffney, Eamonn A},
  journal={Bulletin of Mathematical Biology},
  volume={83},
  pages={1--32},
  year={2021},
  publisher={Springer}
}

@article{kolokolnikov2018pattern,
  title={Pattern formation in a reaction-diffusion system with space-dependent feed rate},
  author={Kolokolnikov, Theodore and Wei, Juncheng},
  journal={SIAM Review},
  volume={60},
  number={3},
  pages={626--645},
  year={2018},
  publisher={SIAM}
}

@article{page2003pattern,
  title={Pattern formation in spatially heterogeneous {T}uring reaction--diffusion models},
  author={Page, K. and Maini, P. K. and Monk, N. A. M.},
  journal={Physica D: Nonlinear Phenomena},
  volume={181},
  number={1-2},
  pages={80--101},
  year={2003},
  publisher={Elsevier}
}

@article{page2005complex,
  title={Complex pattern formation in reaction--diffusion systems with spatially varying parameters},
  author={Page, Karen M and Maini, Philip K and Monk, Nicholas AM},
  journal={Physica D: Nonlinear Phenomena},
  volume={202},
  number={1-2},
  pages={95--115},
  year={2005},
  publisher={Elsevier}
}

@article{iron2001spike,
  title={Spike pinning for the {G}ierer--{M}einhardt model},
  author={Iron, D. and Ward, M. J.},
  journal={Mathematics and computers in simulation},
  volume={55},
  number={4-6},
  pages={419--431},
  year={2001},
  publisher={Elsevier}
}

@article{pelz2026oscillations,
  title={Oscillations in a scalar differential equation coupled to a diffusive field},
  author={Pelz, Merlin and Scheel, Arnd},
  journal={arXiv preprint arXiv:2604.01135},
  year={2026}
}

@article{prigogine1978time,
  title={Time, structure, and fluctuations},
  author={Prigogine, Ilya},
  journal={Science},
  volume={201},
  number={4358},
  pages={777--785},
  year={1978},
  publisher={American Association for the Advancement of Science}
}

@article{matano1983pattern,
  title={Pattern formation in competition-diffusion systems in nonconvex domains},
  author={Matano, Hiroshi and Mimura, Masayasu},
  journal={Publications of the Research Institute for Mathematical Sciences},
  volume={19},
  number={3},
  pages={1049--1079},
  year={1983},
  publisher={Research Institute forMathematical Sciences}
}

@article{kuznetsov2017pattern,
  title={Pattern formation in a reaction-diffusion system of Fitzhugh-Nagumo type before the onset of subcritical Turing bifurcation},
  author={Kuznetsov, Maxim and Kolobov, Andrey and Polezhaev, Andrey},
  journal={Physical Review E},
  volume={95},
  number={5},
  pages={052208},
  year={2017},
  publisher={APS}
}

@article{sanchez2019turing,
  title={Turing--Hopf patterns on growing domains: the torus and the sphere},
  author={S{\'a}nchez-Garduno, Faustino and Krause, Andrew L and Castillo, Jorge A and Padilla, Pablo},
  journal={Journal of theoretical biology},
  volume={481},
  pages={136--150},
  year={2019},
  publisher={Elsevier}
}

@article{wang2022periodic,
  title={Periodic spatial patterning with a single morphogen},
  author={Wang, Sheng and Garcia-Ojalvo, Jordi and Elowitz, Michael B},
  journal={Cell Systems},
  volume={13},
  number={12},
  pages={1033--1047},
  year={2022},
  publisher={Elsevier}
}

@article{fussell2019hybrid,
  title={Hybrid approach to modeling spatial dynamics of systems with generalist predators},
  author={Fussell, Elizabeth F and Krause, Andrew L and Van Gorder, Robert A},
  journal={Journal of theoretical biology},
  volume={462},
  pages={26--47},
  year={2019},
  publisher={Elsevier}
}

\end{document}